# Tasks for Temporal Graph Visualisation

Natalie Kerracher, Jessie Kennedy, and Kevin Chalmers
Institute for Informatics and Digital Innovation
Edinburgh Napier University, Scotland, UK
e-mail: n.kerracher@napier.ac.uk

In [1], we describe the design and development of a task taxonomy for temporal graph visualisation. This paper details the full instantiation of that task taxonomy. Our task taxonomy is based on the Andrienko framework [2], which uses a systematic approach to develop a formal task framework for visual tasks specifically associated with Exploratory Data Analysis. The Andrienko framework is intended to be applicable to all types of data, however, it does not consider relational (graph) data. We therefore extended both their data model and task framework for temporal graph data, and instantiated the extended version to produce a comprehensive list of tasks of interest during exploratory analysis of temporal graph data. As expected, our instantiation of the framework resulted in a very large task list; with more than 144 variations of attribute based tasks alone, it is too large to fit in a standard journal paper, hence we provide the detailed listing in this document.

This paper is organised as follows: in section 1 we give a short overview of the task categories of the taxonomy, a more detailed explanation of which is provided in [1]. A key notion in the Andrienko framework is that of behaviours: in section 2 we briefly summarise the partial and aspectual behaviours of interest when analysing temporal graphs. In section 3 we provide a short guide to the formal notation used in the task definitions of the original framework. Finally, in section 4, we give the complete task listings for both structural and attribute based graph visualisation tasks.

## 1 Overview of task categories

Under the Andrienko framework, there are two components to every task: the target, or unknown information to be obtained, and the constraints, or the known conditions that information needs to fulfil. These targets and constraints distinguish the tasks and determine the shape of the model. We provide an overview of the task categories in Figure 1, which is based on Aigner et al.'s [3] representation of the original Andrienko task model organised into a taxonomy. We have redrawn and extended their figure to include our addition of structural tasks to the taxonomy. A detailed description of the task categories is provided in [1].

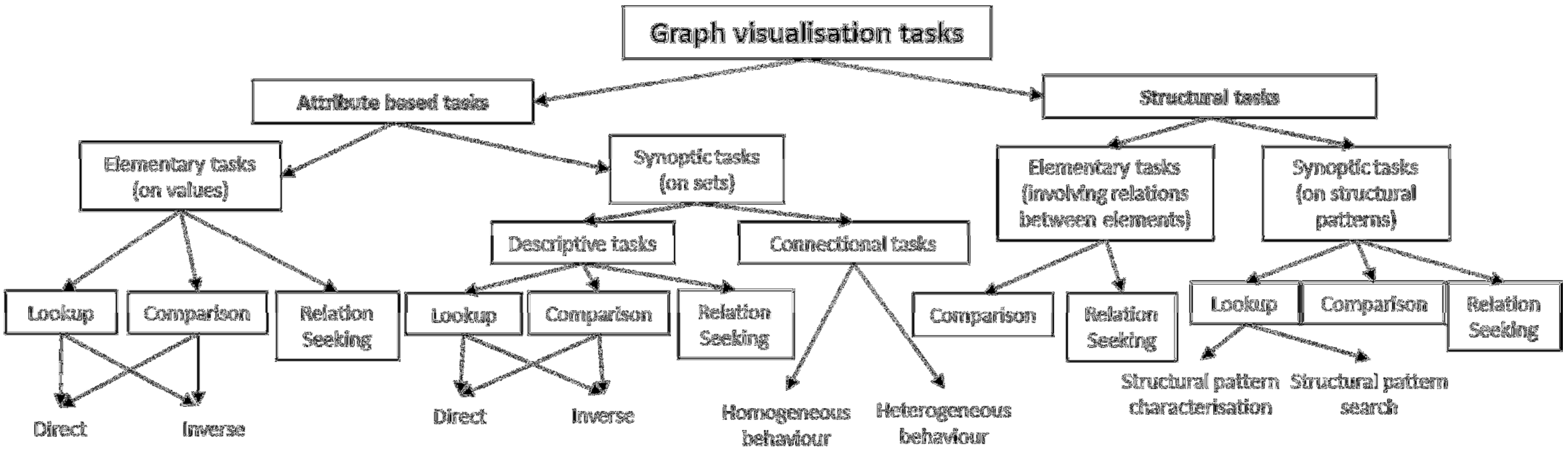

**Figure 1 The task model. Based on Aigner et al.'s drawing [2, p74] of the Andrienko task model organised into a taxonomy, redrawn and extended to include structural tasks for graph visualisation.**

# 2 Behaviours

Behaviours are a key notion in the Andrienko framework: they are representative of real-world phenomena, describing the configurations of sets of attribute values over the independent (referential) component of the data, as determined by the data function (the mapping between the independent and dependent data components), and the relations which exist between the elements of the independent data component e.g. distance and order. A *pattern* results from an observation of a behaviour and provides a descriptive summary of its essential features. Four main types of pattern are distinguished under the Andrienko framework: association, differentiation, arrangement and distribution. Where there are multiple independent components of the data (as in the case of temporal graphs), they distinguish between overall behaviours (which consider all behaviours over the entire dataset) and aspectual behaviours (which consider only certain aspects of the overall behaviour).

In addition to the attribute based behaviours of the Andrienko framework, we introduce the notion of structural behaviours in temporal graphs. We here outline the two partial and two aspectual attribute-based behaviours applicable to temporal graphs, along with the four analogous structural behaviours.

## 2.1 Attribute based behaviours

Partial behaviours:

1. The behaviour of an attribute of a single graph element (a node, edge, or graph object) over time (the whole time period or a subset of time) e.g. a temporal trend in the attribute of a node.
2. The behaviour of an attribute over a set of nodes (or subset of nodes/graph object) at a single time e.g. the distribution of the attribute values over the set.

Aspectual behaviours:

1. The behaviour of the temporal trends (1, above) over the graph (i.e. the distribution of temporal behaviours over the graph).
2. The behaviour over time of the behaviours of attribute values over the set of graph objects (2, above) (i.e. the temporal trend in the distribution of the attribute values over the graph).

## 2.2 Structural behaviours

(1) The behaviour of association relations between two graph objects over time e.g. the pattern of change in connectivity between two nodes over time.

(2) The behaviour, or configuration, of association relations within a set of nodes at a single time e.g. clusters, cliques, motifs etc.

(3) The behaviour of the collection of patterns in (1) i.e. the aggregate pattern of all association relations between pairs of graph objects over time, or the distribution of individual temporal behaviours over the graph.

(4) The behaviour of the configurations of association relations over the set of nodes (i.e. (2)), over time.

# 3 Formal Notation

This section provides a brief summary of the formal notation used to represent variations in tasks in the framework.

## 3.1 Data function applied to temporal graphs

In the case of temporal graphs, we use the following formalism to represent the Andrienko data function which maps a graph element at a particular time point to the corresponding values of the attributes in the data set:

$$\boldsymbol{f}(t, g) = (y_1, y_2, \ldots, y_N)$$

Where:

$t$ represents a time point

$g$ represents a graph element (node, edge, graph object)

$y_1, y_2, \ldots, y_N$ represents the N attributes in the data set

## 3.2 Key to formal notation

| | |
|---|---|
| **Bold** | a specified value (constant) |
| *Italics* | an unknown value (variable) |
| t | a time point |
| T′ | a (sub)set of time points/a time interval |
| g | a graph element (node, edge, graph object) |
| G′, G″ | a (sub)set of graph elements |
| *y* | the value of an unknown characteristic |
| **c** | a specified characteristic |
| **C′** | a subset of characteristics |
| Λ, Ψ, Φ, λ, ψ, ϕ | a relation (e.g. $y_1 \lambda y_2$ can be read as 'the relation between' $y_1$ and $y_2$) |
| $\boldsymbol{\beta}(\boldsymbol{f}(x_1, x_2) \mid x_1 \in \mathbf{G'}, x_2 \in \mathbf{T'})$ | the behaviour $\boldsymbol{\beta}$ of a data function $\boldsymbol{f}$ over the set of graph objects **G′**, and time interval **T′**, where $x_1$ is a graph object in the set of graph objects (**G′**) and $x_2$ is a time point in the time interval (**T′**) |
| $\boldsymbol{\beta}_G\{\boldsymbol{\beta}_T[\boldsymbol{f}(x_1, x_2) \mid x_2 \in \mathbf{T})] \mid x_1 \in \mathbf{G}\}$ <br> $\boldsymbol{\beta}_T\{\boldsymbol{\beta}_G[\boldsymbol{f}(x_1, x_2) \mid x_1 \in \mathbf{G})] \mid x_2 \in \mathbf{T}\}$ | formulae representing the two aspectual behaviours: the behaviour of the temporal behaviours (trends) over the graph (i.e. the distribution of temporal behaviours over the graph), and the behaviour over time of the behaviours (distributions) of attribute values over the set of graph elements (i.e. the temporal trend in the distribution of the attribute values). |
| **P** | a known pattern |
| *p* | An unknown pattern |
| ≈ | 'approximates' |

# 4 Tasks

In this section we list the tasks associated with analysis of temporal graph data. In order to systematically specify all possible permutations of tasks under the framework, we used a series of task matrices when generating the tasks. The comparison and relation seeking matrices can also be found in their complete form at http://www.iidi.napier.ac.uk/c/downloads/downloadid/13377254 for easier reading and printing.

## 4.1 Attribute based tasks

### 4.1.1 Lookup tasks

#### *4.1.1.1 Elementary lookup*

In the temporal graph case, elementary lookup tasks involve the correspondence between a graph element (a node, edge, or graph object) at a particular time point, and its associated attribute value. In *direct lookup*, given a graph element at a given time point, we seek to find the corresponding attribute value. In *inverse lookup*, we seek to find the graph element(s) and/or time point(s) associated with a given attribute value. There are a number of possible variations of inverse lookup, depending on the additional constraints involved: we may specify just the attribute value, or in addition specify either a time point or graph object. These variations are shown in the lookup task matrix (Figure 3).

Additional task variations which are not shown in the matrix, but can be formulated based on the tasks in the matrix, include:

- The case where the attribute value is imprecisely specified, and we allow a set of attribute values e.g. where the value is 'greater than 50' or in the set {red, green, blue}: ? $t, g$: $\boldsymbol{f}(t, g) \in \mathbf{C'}$
- Specifying a subset of graph elements or time points (e.g. an interval) (in place of a single graph element or time point) as an additional constraint e.g.
    - Find the time(s) at which *any graph object in the specified* subset have the given attribute value: ? $t$: $\boldsymbol{f}(t, \mathbf{G'}) = \mathbf{c}$
    - Find the graph object(s) which have the given attribute value *at any time during the given time interval* ? $g$: $\boldsymbol{f}(\mathbf{T'}, g) = \mathbf{c}$
- Where the values of either time or graph are of no importance, we allow the *whole* set of time points or graph elements to be specified:
    - Find the time(s) at which *any* graph object had the given attribute value: ? $t$: $\boldsymbol{f}(t, \mathbf{G}) = \mathbf{c}$
    - Find the graph object(s) which had the given attribute value at any time: ? $g$: $\boldsymbol{f}(\mathbf{T}, g) = \mathbf{c}$

#### *4.1.1.2 Synoptic lookup*

Behaviour characterisation involves finding the pattern which approximates the behaviour of an attribute over a reference set (or subset). Pattern search is the opposite: given a pattern, we find the subset of references over which the behaviour corresponds to the specified pattern. In the temporal graph case, these tasks involve the aspectual and partial behaviours described in section 2, and the corresponding graph and temporal references. This results in three task variations, depending on the referential components involved. We outline these in quadrants 2-4 of Figure 2. Further variations in the pattern search task depend on which referential components are specified, and these are detailed in the full lookup task matrix (Figure 3).

**Figure 2 Quadrant-level overview of the lookup task matrix**

| | Graph Elements (nodes, edges, graph objects) | Graph subsets |
|---|---|---|
| **Time Points** | **Q1 Elementary**<br><br>**Task components:**<br>Referrers are graph elements and time points; characteristics are attribute values.<br><br>**Direct lookup**<br>$?y\colon \boldsymbol{f}(\mathbf{t}, \mathbf{g}) = y$<br>Involves finding the attribute value of a given graph element at a given time point.<br><br>**Inverse lookup**<br>$?\, t, g\colon \boldsymbol{f}(t, g) = \mathbf{c}$<br>Involves finding the graph element(s)/time point(s) associated with a given attribute value | **Q2 Synoptic**<br><br>**Task components:**<br>The referential component involves the whole graph (or a subset of the graph) and a single time point; behaviour is that of an attribute over the graph (at a single time).<br><br>**Behaviour characterisation**<br>$?p\colon \boldsymbol{\beta}(\boldsymbol{f}(x_1, x_2) \mid x_1 \in \mathbf{G}, x_2 = \mathbf{t}) \approx p$<br>Involves finding the pattern which approximates the behaviour of an attribute over the graph (or a specified subset of the graph) at the given time point<br><br>**Pattern search**<br>$?G, t\colon \boldsymbol{\beta}(\boldsymbol{f}(x_1, x_2) \mid x_1 \in G, x_2 = t) \approx \mathbf{P}$<br>Involves finding the time point(s) and/or subset(s) of graph elements over which a given pattern of attributes occur. |
| **Time Intervals** | **Q3 Synoptic**<br><br>**Task components:**<br>The referential component involves the whole time period (or a time interval) and a single graph element; behaviour is that of an attribute of a single graph element over time.<br><br>**Behaviour characterisation**<br>$?p\colon \boldsymbol{\beta}(\boldsymbol{f}(x_1, x_2) \mid x_1 = \mathbf{g}, x_2 \in \mathbf{T}) \approx p$<br>Involves finding the pattern which approximates the behaviour of an attribute of a given graph element over the whole time period (or a specified time interval)<br><br>**Pattern search**<br>$?g, T\colon \boldsymbol{\beta}(\boldsymbol{f}(x_1, x_2) \mid x_1 = g, x_2 \in T) \approx \mathbf{P}$<br>Involves finding the graph element(s) and/or time interval(s) over which a given pattern of attributes occurs. | **Q4 Synoptic**<br><br>**Task components:**<br>The referential component involves the whole time period (or a time interval) and the whole graph (or a subset of the graph); behaviour is either of the two aspectual behaviours: the distribution of temporal trends over the graph or the distributions of an attribute over the graph, over time.<br><br>**Behaviour characterisation**<br>Involves finding the pattern that approximates the aspectual behaviours:<br>$?p\colon \boldsymbol{\beta}_G\{\boldsymbol{\beta}_T[\boldsymbol{f}(x_1, x_2) \mid x_2 \in \mathbf{T})] \mid x_1 \in \mathbf{G}\} \approx p$<br>the behaviour of the temporal behaviours (trends) over the graph (i.e. the distribution of temporal behaviours over the graph)<br><br>or<br>$?p\colon \boldsymbol{\beta}_T\{\boldsymbol{\beta}_G[\boldsymbol{f}(x_1, x_2) \mid x_1 \in \mathbf{G})] \mid x_2 \in \mathbf{T}\} \approx p$<br>the behaviour over time of the behaviours (distributions) of attribute values over the set of graph objects (i.e. the temporal trend in the distribution of the attribute values); in both cases we may be interested in the behaviour associated with a given subset of the time period or the graph.<br><br>**Pattern search**<br>$?T, G\colon \boldsymbol{\beta}_G\{\boldsymbol{\beta}_T[\boldsymbol{f}(x_1, x_2) \mid x_2 \in T)] \mid x_1 \in G\} \approx \mathbf{P}$<br>or<br>$?G, T\colon \boldsymbol{\beta}_T\{\boldsymbol{\beta}_G[\boldsymbol{f}(x_1, x_2) \mid x_1 \in G)] \mid x_2 \in T\} \approx \mathbf{P}$<br>Involves finding the subset(s) of time and/or graph elements over which a (sub)pattern of an aspectual behaviour occurs. |

Figure 3 Lookup task matrix

| | | Graph Elements | | Graph subsets | |
|---|---|---|---|---|---|
| | | **Constraint** | **Target** | **Constraint** | **Target** |
| **Time point** | **Target** | **Direct look up** given a graph object and time, find the attribute value<br><br>$?y\colon \boldsymbol{f}(\mathbf{t}, \mathbf{g}) = y$ | **Inverse lookup** given an attribute value and a time point, find the graph object(s) which have this value<br><br>$?\ g\colon \boldsymbol{f}(\mathbf{t}, g) = \mathbf{c}$ | **Behaviour characterisation**<br>Find the pattern that approximates (i.e. characterise) the behaviour of an attribute over the graph (or a subset of the graph) at the given time point<br><br>$?p\colon \boldsymbol{\beta}(\boldsymbol{f}(x_1, x_2) \mid x_1 \in \mathbf{G}, x_2 = \mathbf{t}) \approx p$ | **Pattern search** find the subset(s) of the graph over which a particular pattern of attribute values occurs, at the given time point<br><br>$?G\colon \boldsymbol{\beta}(\boldsymbol{f}(x_1, x_2) \mid x_1 \in G, x_2 = \mathbf{t}) \approx \mathbf{P}$ |
| | **Constraint** | **Inverse look up** given a graph object and attribute value, find the time point(s) at which it occurs<br><br>$?\ t\colon \boldsymbol{f}(t, \mathbf{g}) = \mathbf{c}$ | **Inverse lookup** given an attribute value, find the graph object(s), and the time point(s), at which the value occurs<br><br>$?\ t, g\colon \boldsymbol{f}(t, g) = \mathbf{c}$ | **Pattern search** find the time point(s) at which a particular pattern of attributes over the graph occurs<br><br>$?\ t\colon \boldsymbol{\beta}(\boldsymbol{f}(x_1, x_2) \mid x_1 \in \mathbf{G}, x_2 = t) \approx \mathbf{P}$ | **Pattern search** find the time point(s) and subset(s) of the graph over which a particular pattern of attribute values occurs<br>$?G, t\colon \boldsymbol{\beta}(\boldsymbol{f}(x_1, x_2) \mid x_1 \in G, x_2 = t) \approx \mathbf{P}$<br><br>e.g. find (connected) subsets of the graph which have very similar attribute values, and the time points at which they occur |
| **Time interval** | **Target** | **Behaviour characterisation** characterise the behaviour of a attribute of a single node over time.<br><br>$?p\colon \boldsymbol{\beta}(\boldsymbol{f}(x_1, x_2) \mid x_1 = \mathbf{g}, x_2 \in \mathbf{T}) \approx p$ | **Pattern search** find the node(s) over which a particular pattern of attribute values occurs, over the given time interval.<br><br>$?g\colon \boldsymbol{\beta}(\boldsymbol{f}(x_1, x_2) \mid x_1 = g, x_2 \in \mathbf{T}) \approx \mathbf{P}$ | **Behaviour characterisation**<br>(i) characterise the behaviour of the temporal trends over the graph (i.e. the distribution of temporal behaviours over the graph)<br>$?p\colon \boldsymbol{\beta}_G\{\boldsymbol{\beta}_T[\boldsymbol{f}(x_1, x_2) \mid x_2 \in \mathbf{T})] \mid x_1 \in \mathbf{G}\} \approx p$<br><br>(ii) characterise the behaviour of the attribute values over the graph, over time<br>$?p\colon \boldsymbol{\beta}_T\{\boldsymbol{\beta}_G[\boldsymbol{f}(x_1, x_2) \mid x_1 \in \mathbf{G})] \mid x_2 \in \mathbf{T}\} \approx p$ | **Pattern search**<br>(i)Find the subset(s) of graph elements over which a given pattern in the collection of temporal trends occurs, over the given time interval<br>$?\ G\colon \boldsymbol{\beta}_G\{\boldsymbol{\beta}_T[\boldsymbol{f}(x_1, x_2) \mid x_2 \in \mathbf{T})] \mid x_1 \in G\} \approx \mathbf{P}$<br><br>(ii) find the subset(s) of the graph over which a given (temporal) pattern in the pattern of attribute values over the graph occurs<br>$?G\colon \boldsymbol{\beta}_T\{\boldsymbol{\beta}_G[\boldsymbol{f}(x_1, x_2) \mid x_1 \in G)] \mid x_2 \in \mathbf{T}\} \approx \mathbf{P}$ |
| | **Constraint** | **Pattern search** find the time interval over which a given pattern of attribute values occurs for a given node.<br><br>$?T\colon \boldsymbol{\beta}(\boldsymbol{f}(x_1, x_2) \mid x_1 = \mathbf{g}, x_2 \in T) \approx \mathbf{P}$ | **Pattern search** find the node(s) and time interval(s) over which the specified pattern of attribute values occurs<br><br>$?g, T\colon \boldsymbol{\beta}(\boldsymbol{f}(x_1, x_2) \mid x_1 = g, x_2 \in T) \approx \mathbf{P}$ | **Pattern search**<br>(i)Find the time interval(s) over which a given pattern in the collection of temporal trends occurs<br>$?\ T\colon \boldsymbol{\beta}_G\{\boldsymbol{\beta}_T[\boldsymbol{f}(x_1, x_2) \mid x_2 \in T)] \mid x_1 \in \mathbf{G}\} \approx \mathbf{P}$<br><br>(ii) find the time interval(s) over which a given (temporal) pattern in the pattern of attribute values over the graph occurs<br>$?T\colon \boldsymbol{\beta}_T\{\boldsymbol{\beta}_G[\boldsymbol{f}(x_1, x_2) \mid x_1 \in \mathbf{G})] \mid x_2 \in T\} \approx \mathbf{P}$ | **Pattern search**<br>(i) Find the subset(s) of graph elements and time interval(s) over which a given pattern in the collection of temporal trends occurs<br>$?T, G\colon \boldsymbol{\beta}_G\{\boldsymbol{\beta}_T[\boldsymbol{f}(x_1, x_2) \mid x_2 \in T)] \mid x_1 \in G\} \approx \mathbf{P}$<br><br>(ii) Find the time interval(s) and subset(s) of the graph over which a given (temporal) pattern in the pattern of attribute values over the graph occurs<br>$?G, T\colon \boldsymbol{\beta}_T\{\boldsymbol{\beta}_G[\boldsymbol{f}(x_1, x_2) \mid x_1 \in G)] \mid x_2 \in T\} \approx p$ |

### 4.1.2 Comparison

Comparison tasks are compound tasks, which consist of lookup tasks to find the elements to be compared, and comparison of these elements to find the relation between them. Direct and inverse comparison are distinguished based on the lookup tasks and resulting elements involved in the comparison subtask. In the elementary case, in direct comparison, we use direct lookup and compare the found attribute values; in inverse comparison, we use inverse lookup and compare references (time points and/or graph elements). In the synoptic case, direct comparison involves behaviour characterisation subtasks and comparison of patterns, while inverse comparison involves pattern search subtasks and comparison of the associated graph subsets and/or time intervals.

#### 4.1.2.1 Direct comparison

In both the elementary and synoptic direct comparison case, four subtasks are distinguished in the Andrienko framework, depending on whether:

- One of the attribute values/patterns involved in the comparison is *specified* or two lookup/behaviour characterisation subtasks are required
- the *attributes* involved in each lookup/behaviour characterisation task are the same or different
- the *references* involved in each lookup task are the same or different

As temporal graphs have two referrers, "the same reference" implies the same graph element or graph subset at the same point in time or ($\mathbf{t}_1=\mathbf{t}_2$, $\mathbf{g}_1=\mathbf{g}_2$; we use just **t** and **g** to indicate this in the task listings); there are three possible variations of what could be meant by "different references" (note that the same applies to subsets of the graph and time intervals):

a. The same graph object at different time points ($\mathbf{t}_1\neq\mathbf{t}_2$, $\mathbf{g}_1=\mathbf{g}_2$)

b. Different graph objects at the same time point ($\mathbf{t}_1=\mathbf{t}_2$, $\mathbf{g}_1\neq\mathbf{g}_2$)

c. Two different graph objects at two different time points ($\mathbf{t}_1\neq\mathbf{t}_2$, $\mathbf{g}_1\neq\mathbf{g}_2$)

Task a. is the typical temporal graph scenario: we refer to this type of task as an *evolutionary* task, as we are interested in how the properties of a graph element have changed or evolved between time points. We refer to tasks b and c as *contextual* tasks, as we often carry out such tasks in order to put the properties of one graph object in the context of another. Task b is also applicable to static graphs, as this is equivalent to considering the graph at a single time point.

Note that **we do not show the variations of tasks involving the same/different attributes in the task matrix,** but all tasks (with the exception of direct comparisons involving the same time point/interval and graph element/subset) could potentially be formulated to consider comparison involving the same attributes or two different attributes in the lookup subtask.

#### 4.1.2.2 Inverse comparison

Three variations of the inverse comparison task are identified in the Andrienko framework based on:

- Whether two inverse lookup subtasks are involved, or one of the references (time point/graph element) or reference subsets (time interval/graph subset) is specified.

- Whether the attribute involved in each subtask is the same or two different attributes are considered

As noted above, we do not show variations of tasks involving the same/different attributes, but these can potentially be formulated for each task. The large number of tasks in the task matrix are derived from the permutations of references (time point/interval, graph element/subset) and to what degree the references in the subtasks are specified.

We summarise the comparison tasks at the quadrant level, based on the references involved, in Figure 4.

Figure 4 Quadrant-level overview of the comparison task matrix

| | Graph Elements (nodes, edges, graph objects) | Graph subsets |
|---|---|---|
| **Time points** | **Q1 Elementary**<br><br>**Direct comparison**<br>? $y_1, y_2, \lambda$: $\boldsymbol{f_1}(\mathbf{t_1}, \mathbf{g_1}) = y_1$; $\boldsymbol{f_2}(\mathbf{t_2}, \mathbf{g_2}) = y_2$; $y_1 \lambda\, y_2$<br>- of attribute values associated with a given graph element at a given time (the attribute involved in the lookup tasks may be the same or different, hence the data functions $\boldsymbol{f_1}(\mathbf{x})$ and $\boldsymbol{f_2}(\mathbf{x})$).<br>*Relations:*<br>• *between attribute values are domain dependent.*<br><br>**Inverse comparison**<br>? $t_1, t_2, g_1, g_2, \lambda$: $\boldsymbol{f}(t_1, g_1) \in \mathbf{C'}$; $\boldsymbol{f}(t_2, g_2) \in \mathbf{C''}$; $(t_1, g_1)\, \lambda (t_2, g_2)$<br><br>- of two graph elements and/or two time points associated with given attribute values<br><br>*Relations:*<br>• *between graph elements: equality (same/different element); set relations (between the sets of elements belonging to graph objects); equality of configuration (in graph objects); association (between nodes/graph objects, at a single time point only);*<br>• *between two time points: happens before(/after), happens at the same time* [4]. | **Q2 Synoptic**<br><br>**Direct comparison**<br>? $p_1, p_2, \lambda$:<br>$\boldsymbol{\beta}(\boldsymbol{f}(x_1, x_2) \mid x_1 \in \mathbf{G'}, x_2 = \mathbf{t_1}) \approx p_1$;<br>$\boldsymbol{\beta}(\boldsymbol{f}(x_1, x_2) \mid x_1 \in \mathbf{G''}, x_2 = \mathbf{t_2}) \approx p_2$;<br>$p_1 \lambda\, p_2$<br>**–** of two patterns of an attribute(s)[1] over the graph (or a subset of the graph elements) at given time point(s)<br>*Relations:*<br>• *between patterns: same(similar)/different/opposite*[2]<br>**Inverse comparison**<br>? $G', G'', t_1, t_2, \lambda, \psi$:<br>$\boldsymbol{\beta}(\boldsymbol{f}(x_1, x_2) \mid x_1 \in G', x_2 = t_1) \approx \mathbf{P_1}$;<br>$\boldsymbol{\beta}(\boldsymbol{f}(x_1, x_2) \mid x_1 \in G'', x_2 = t_2) \approx \mathbf{P_2}$;<br>$(G', t_1)\, \lambda\, (G'', t_2)$;<br>$t_1 \psi\, t_2$<br><br>- of the time points at which the given patterns occur<br>- of the graph subsets over which a given pattern occurs;<br>- comparison of both time points and graph subsets.<br><br>*Relations:*<br>• *between two time points: happens before(/after), happens at the same time* [4];<br>• *between two graph subsets: equality (same/different subset); set relations (between the sets of nodes/edges belonging to the subset); equality of configuration (of the subset); association (between nodes/graph objects, at a single time point only).* |
| **Time intervals** | **Q3 Synoptic**<br><br>**Direct comparison**<br>? $p_1, p_2, \lambda$: $\boldsymbol{\beta}(\boldsymbol{f}(x_1, x_2) \mid x_1 = \mathbf{g_1}, x_2 \in \mathbf{T'}) \approx p_1$; $\boldsymbol{\beta}(\boldsymbol{f}(x_1, x_2) \mid x_1 = \mathbf{g_2}, x_2 \in \mathbf{T''}) \approx p_2$; $p_1 \lambda\, p_2$<br>- of two (temporal) patterns associated with an attribute(s)[Error! Bookmark not defined.] of given graph element(s) | **Q4 Synoptic**<br><br>**Direct comparison**<br>? $p_1, p_2, \lambda$: $\boldsymbol{\beta_G}\{\boldsymbol{\beta_T}[\boldsymbol{f}(x_1, x_2) \mid x_2 \in \mathbf{T'})] \mid x_1 \in \mathbf{G'}\} \approx p_1$; $\boldsymbol{\beta_G}\{\boldsymbol{\beta_T}[\boldsymbol{f}(x_1, x_2) \mid x_2 \in \mathbf{T''})] \mid x_1 \in \mathbf{G''}\} \approx p_2$; $p_1 \lambda\, p_2$ (comparison of patterns of distributions of temporal trends over the graph)<br>or |

[1] i.e. each pattern may correspond to a different attribute

[2] In descriptive synoptic tasks (in connectional synoptic tasks, patterns of “mutual” behaviours include correlation, dependency, and structural connection.

| | Graph Elements (nodes, edges, graph objects) | Graph subsets |
|---|---|---|
| | over the whole time period (or a specified time interval)<br>*Relations:*<br>• *between patterns: same (similar)/different/opposite*<br><br>**Inverse comparison**<br>$? g_1, g_2, T', T'', \lambda, \psi$: $\boldsymbol{\beta}(\boldsymbol{f}(x_1, x_2) \mid x_1 = g_1, x_2 \in T') \approx \mathbf{P_1}$; $\boldsymbol{\beta}(\boldsymbol{f}(x_1, x_2) \mid x_1 = g_2, x_2 \in T'') \approx \mathbf{P_2}$; $g_1 \lambda g_2$; $T' \psi T''$<br>– of the time intervals over which given patterns occur; of the graph elements associated with a given pattern; comparison of both time intervals and graph elements<br>*Relations:*<br>• *between two graph elements: equality (same/different; set relations between the sets of elements belonging to graph objects);*<br>• *between time intervals: happens before(/after), happens at the same time; between two intervals, or an instant and an interval: happens before(/after), starts, finishes, happens during; between intervals only: overlaps, meets* [4]. | $? p_1, p_2, \lambda$: $\boldsymbol{\beta}_T\{\boldsymbol{\beta}_G[\boldsymbol{f}(x_1, x_2) \mid x_1 \in \mathbf{G'})] \mid x_2 \in \mathbf{T'}\} \approx p_1$; $\boldsymbol{\beta}_T\{\boldsymbol{\beta}_G[\boldsymbol{f}(x_1, x_2) \mid x_1 \in \mathbf{G''})] \mid x_2 \in \mathbf{T''}\} \approx p_2$; $p_1 \lambda p_2$<br>(comparison of patterns of distributions of an attribute over the graph, over time)<br><br>– of two patterns associated with a given subset of time and/or subset of graph elements. The patterns may reflect either of the two aspectual behaviours (the distribution of temporal trends over the graph or the distributions of an attribute over the graph, over time)<br>*Relations*<br>• *between patterns: same (similar)/different/opposite*<br><br>**Inverse comparison**<br>$? G', G'', T', T'', \lambda, \psi$: $\boldsymbol{\beta}_G\{\boldsymbol{\beta}_T[\boldsymbol{f}(x_1, x_2) \mid x_2 \in T')] \mid x_1 \in G'\} \approx \mathbf{P_1}$; $\boldsymbol{\beta}_G\{\boldsymbol{\beta}_T[\boldsymbol{f}(x_1, x_2) \mid x_2 \in T'')] \mid x_1 \in G''\} \approx \mathbf{P_2}$; $T' \lambda T''$; $G' \psi G''$;<br><br>or<br>$? G', G'', T', T'', \lambda, \psi$: $\boldsymbol{\beta}_T\{\boldsymbol{\beta}_G[\boldsymbol{f}(x_1, x_2) \mid x_1 \in G')] \mid x_2 \in T'\} \approx \mathbf{P_1}$; $\boldsymbol{\beta}_T\{\boldsymbol{\beta}_G[\boldsymbol{f}(x_1, x_2) \mid x_1 \in G'')] \mid x_2 \in T''\} \approx \mathbf{P_2}$; $T' \lambda T''$; $G' \psi G''$;<br><br>– of the time intervals and/or subsets of graph elements associated with a given aspectual (sub)pattern<br>*Relations:*<br>• *between two graph subsets: equality, set relations*<br>• *between time intervals: happens before(/after), happens at the same time; between two intervals, or an instant and an interval: happens before(/after), starts, finishes, happens during; between intervals only: overlaps, meets* [4]. |

Notes on comparison task matrix:

- Due to issues of space on the printed page, we here show each quadrant of the comparison task matrix separately. The compiled task matrix can be found at http://www.iidi.napier.ac.uk/c/downloads/downloadid/13377254.
- In the following tasks we use ($G'$, $t_1$) to specify a graph subset at a given time (as opposed to just $G'$). This is due to the nature of the graph referrer: as association relations in the graph referrer may change over time, a graph object at $t_1$ may be quite different from "the same" graph object at $t_2$.
- Where both graph elements/subsets and/or both time points/intervals are unspecified, we can add an additional constraint to the task i.e. that the components in question have a specified relation between them e.g. in the case of the graph referrer, that they are the same, connected, a certain distance from one another etc. or in the case of time that they are the same, overlapping, a given distance from one another etc . Where we restrict graph elements/subsets to being the same, and the temporal component is different, these become evolutionary tasks e.g. compare the time intervals over which two patterns occur over two time intervals for the same graph object:
  $?g, T', T'', \lambda, \psi$: $\boldsymbol{\beta}(\boldsymbol{f}(x_1, x_2) \mid x_1 = g, x_2 \in T') \approx \mathbf{P_1}$; $\boldsymbol{\beta}(\boldsymbol{f}(x_1, x_2) \mid x_1 = g, x_2 \in T'') \approx \mathbf{P_2}$; $T' \psi T''$

Figure 5 Comparison task matrix, quadrant 1: considers comparisons involving graph elements (nodes, edges, graph objects) and time points (i.e. the elementary comparison tasks)

| | | | Graph elements (nodes, edges, graph objects) | | | |
|---|---|---|---|---|---|---|
| | | | Both constraints | | One element specified | Neither element specified |
| | | | Single/same element | Two different elements | | |
| Time points | Both constraints | Same time | **Direct comparison** Compare the values of *different* attributes for a given node at a given time point.<br><br>? $y_1, y_2, \lambda$:<br>$\boldsymbol{f_1}(\mathbf{t}, \mathbf{g}) = y_1$; $\boldsymbol{f_2}(\mathbf{t}, \mathbf{g}) = y_2$;<br>$y_1 \lambda\, y_2$ | **Direct comparison** Compare the attribute values associated with two different nodes at the same time point.<br>? $y_1, y_2, \lambda$:<br>$\boldsymbol{f}(\mathbf{t}, \mathbf{g_1}) = y_1$; $\boldsymbol{f}(\mathbf{t}, \mathbf{g_2}) = y_2$;<br>$y_1 \lambda\, y_2$ | **Inverse comparison** This task reduces to comparison with a specified reference[i] . Find and compare with a given node, the node(s) associated with the given attribute value at the given time.<br><br>? $g_2, \lambda$:<br>$\boldsymbol{f}(\mathbf{t}, g_2) \in \mathbf{C'}$;<br>$(\mathbf{t}, \mathbf{g_1})\ \lambda (\mathbf{t}, g_2)$ | **Inverse comparison** Find and compare the nodes associated with two different attribute values at the given time<br><br>? $g_1, g_2, \lambda$:<br>$\boldsymbol{f}(\mathbf{t}, g_1) \in \mathbf{C'}$; $\boldsymbol{f}(\mathbf{t}, g_2) \in \mathbf{C''}$;<br>$(\mathbf{t}, g_1)\ \lambda (\mathbf{t}, g_2)$ |
| | | Different times | **Direct comparison** Compare the attribute values associated with a single node at two different times.<br><br>? $y_1, y_2, \lambda$:<br>$\boldsymbol{f}(\mathbf{t_1}, \mathbf{g_1}) = y_1$; $\boldsymbol{f}(\mathbf{t_2}, \mathbf{g_2}) = y_2$;<br>$y_1 \lambda\, y_2$ | **Direct comparison** Compare the attribute values associated with two different nodes at two different times.<br><br>? $y_1, y_2, \lambda$:<br>$\boldsymbol{f}(\mathbf{t_1}, \mathbf{g_1}) = y_1$; $\boldsymbol{f}(\mathbf{t_2}, \mathbf{g_2}) = y_2$;<br>$y_1 \lambda\, y_2$ | **Inverse comparison** As above but involving two different time points[ii]. Find and compare with a given node, the node(s) associated with the given attribute value at the given times.<br><br>? $g_2, \lambda$:<br>$\boldsymbol{f}(\mathbf{t_2}, g_2) \in \mathbf{C'}$;<br>$(\mathbf{t_1}, \mathbf{g_1})\ \lambda\ (\mathbf{t_2}, g_2)$ | **Inverse comparison** As above, but involving two different time points. Find and compare the nodes associated with two different attribute values at the given times<br><br>? $g_1, g_2, \lambda$:<br>$\boldsymbol{f}(\mathbf{t_1}, g_1) \in \mathbf{C'}$; $\boldsymbol{f}(\mathbf{t_2}, g_2) \in \mathbf{C''}$;<br>$(\mathbf{t_1}, g_1)\ \lambda (\mathbf{t_2}, g_2)$ |
| | One time point specified | | **Inverse comparison** This task reduces to comparison with a specified reference[iii]. Find the time point(s) associated with the given attribute value for the given node, and compare it with a given time point.<br><br>? $t_2, \lambda$:<br>$\boldsymbol{f}(t_2, \mathbf{g}) \in \mathbf{C'}$;<br>$\mathbf{t_1}\ \lambda\ t_2$ | **Inverse comparison** As left, this task reduces to comparison with a specified reference[iv].<br>? $t_2, \lambda$:<br>$\boldsymbol{f}(t_2, \mathbf{g}) \in \mathbf{C'}$;<br>$\mathbf{t_1}\ \lambda\ t_2$ | **Inverse comparison**<br>Either:<br>A task reduced to comparison with a specified reference[v]. Find the node(s) and time point(s) at which it has a given attribute value, and compare this with a given node at a given time point. | **Inverse comparison** Find the node(s) having a specified attribute value at a given time, and the node(s) and time point(s) having a given attribute value, and compare the nodes and time points.<br><br>? $t_2, g_1, g_2, \lambda$:<br>$\boldsymbol{f}(\mathbf{t_1}, g_1) \in \mathbf{C'}$; $\boldsymbol{f}(t_2, g_2) \in \mathbf{C''}$;<br>$(\mathbf{t_1}, g_1)\ \lambda (t_2, g_2)$ |

| | Graph elements (nodes, edges, graph objects) | | | |
|---|---|---|---|---|
| | Both constraints | | One element specified | Neither element specified |
| | Single/same element | Two different elements | | |
| | | | ? $t_2$, $g_2$, $\lambda$, $\Psi$:<br>$\boldsymbol{f}(t_2, g_2) \in \mathbf{C'}$;<br>$(\mathbf{t_1}, \mathbf{g_1})\ \lambda (t_2, g_2)$;<br>$\mathbf{t_1}\ \Psi\ t_2$<br><br>OR<br><br>Find the time point at which a given node has a given attribute value, and the node which has a given attribute value at a given time, and compare the nodes and time points.<br><br>? $t_1$, $g_2$, $\lambda$, $\Psi$:<br>$\boldsymbol{f}(t_1, \boldsymbol{g_1}) \in \mathbf{C'}$; $\boldsymbol{f}(\mathbf{t_2}, g_2) \in \mathbf{C''}$;<br>$(t_1, \boldsymbol{g_1})\ \lambda (\mathbf{t_2}, g_2)$;<br>$t_1\ \Psi\ \mathbf{t_2}$ | |
| Neither time point specified | **Inverse comparison** Find and compare the times at which the given node had the given attribute values.<br><br>? $t_1$, $t_2$, $\lambda$:<br>$\boldsymbol{f}(t_1, \mathbf{g}) \in \mathbf{C'}$; $\boldsymbol{f}(t_2, \mathbf{g}) \in \mathbf{C''}$;<br>$t_1\ \lambda\ t_2$ | **Inverse comparison** Find and compare the times at which two given nodes had the given attribute values.<br><br>? $t_1$, $t_2$, $\lambda$:<br>$\boldsymbol{f}(t_1, \mathbf{g_1}) \in \mathbf{C'}$; $\boldsymbol{f}(t_2, \mathbf{g_2}) \in \mathbf{C''}$;<br>$t_1\ \lambda\ t_2$ | **Inverse comparison** Find the time point(s) at which a given node had a given attribute value, and the time point(s) and node(s) having a second given attribute value, and compare the nodes and time points.<br><br>? $t_1$, $t_2$, $g_2$, $\lambda$:<br>$\boldsymbol{f}(t_1, \mathbf{g_1}) \in \mathbf{C'}$; $\boldsymbol{f}(t_2, g_2) \in \mathbf{C''}$;<br>$(t_1, \mathbf{g_1})\ \lambda (t_2, g_2)$ | **Inverse comparison** Find the time points and nodes associated with two given attribute values and compare them.<br><br>? $t_1$, $t_2$, $g_1$, $g_2$, $\lambda$:<br>$\boldsymbol{f}(t_1, g_1) \in \mathbf{C'}$; $\boldsymbol{f}(t_2, g_2) \in \mathbf{C''}$;<br>$(t_1, g_1)\ \lambda (t_2, g_2)$ |

**Figure 6 Comparison quadrant 2: considers comparisons involving the behaviour of an attribute over the graph (or a graph subset)**

| | | | Graph subsets | | | |
|---|---|---|---|---|---|---|
| | | | Both constraints | | One constraint, one target | Both are targets |
| | | | Same subset | Different subsets | | |
| Time points | Both constraints | Same time | **Direct comparison** of the attribute patterns *of two different attributes* over the same subset of the graph at the same time point.<br><br>? $p_1, p_2, \lambda$:<br>$\beta(f_1(x_1, x_2) \mid x_1 \in G', x_2 = t) \approx p_1$;<br>$\beta(f_2(x_1, x_2) \mid x_1 \in G', x_2 = {}_2) \approx p_2$;<br>$p_1 \lambda\, p_2$ | **Direct comparison** of the attribute patterns over two different subsets of the graph at the same time point.<br><br>? $p_1, p_2, \lambda$:<br>$\beta(f(x_1, x_2) \mid x_1 \in G', x_2 = t) \approx p_1$;<br>$\beta(f(x_1, x_2) \mid x_1 \in G'', x_2 = t) \approx p_2$;<br>$p_1 \lambda\, p_2$ | **Inverse comparison** of a given graph subset with the graph subset associated with a given pattern at a given time[vi].<br><br>? $G', \lambda$:<br>$\beta(f(x_1, x_2) \mid x_1 \in G', x_2 = t) \approx P$;<br>$G', t) \lambda (G'', t)$ | **Inverse comparison** of two graph subsets associated with two given patterns at the same specified time.<br><br>? $G', G'', \lambda$:<br>$\beta(f(x_1, x_2) \mid x_1 \in G', x_2 = t) \approx P_1$;<br>$\beta(f(x_1, x_2) \mid x_1 \in G'', x_2 = t) \approx P_2$;<br>$(G', t) \lambda (G'', t)$ |
| | | Different times | **Direct comparison** of the attribute patterns over the same subset of the graph at two different time points.<br><br>? $p_1, p_2, \lambda$:<br>$\beta(f(x_1, x_2) \mid x_1 \in G', x_2 = t_1) \approx p_1$;<br>$\beta(f(x_1, x_2) \mid x_1 \in G', x_2 = t_2) \approx p_2$;<br>$p_1 \lambda\, p_2$ | **Direct comparison** of the attribute patterns over two different subsets of the graph at two different time points.<br><br>? $p_1, p_2, \lambda$:<br>$\beta(f(x_1, x_2) \mid x_1 \in G', x_2 = t_1) \approx p_1$;<br>$\beta(f(x_1, x_2) \mid x_1 \in G'', x_2 = t_2) \approx p_2$;<br>$p_1 \lambda\, p_2$ | **Inverse comparison** as above, but the specified subset of graph elements is associated with a different time point:<br><br>? $G', \lambda$:<br>$\beta(f(x_1, x_2) \mid x_1 \in G', x_2 = t_1) \approx P$;<br>$(G', t_1) \lambda (G'', t_2)$; | **Inverse comparison** of two graph subsets associated with two given patterns at two different, specified time points.<br><br>? $G', G'', \lambda$:<br>$\beta(f(x_1, x_2) \mid x_1 \in G', x_2 = t_1) \approx P_1$;<br>$\beta(f(x_1, x_2) \mid x_1 \in G'', x_2 = t_2) \approx P_2$;<br>$(G', t_1) \lambda (G'', t_2)$ |
| | One constraint, one target | | **Inverse comparison** of the time point associated with a given pattern over a given subset of the graph, with a given time point[3]. | **Inverse comparison**, as left[4] | **Inverse comparison** of a given graph subset at a given time with the graph subset associated with a given pattern, *and* comparison of a given time point with the time point also associated with the given | **Inverse comparison** of the graph objects associated with two patterns, one of them occurring at a given time, *and* comparison of the given time point with the unknown time point at which the second pattern occurs. |

[3] Reduced from:? $t_2, \lambda$: $\beta(f(x_1, x_2) \mid x_1 \in G', x_2 = t_1) \approx P_1$; $\beta(f(x_1, x_2) \mid x_1 \in G', x_2 = t_2) \approx P_2$; $t_1 \lambda\, t_2$ i.e. all information (the graph subset, time point and pattern) is known in the first lookup subtask.

[4] NB in this case, the formula from which this task is reduced would have involved two different graph subsets.

| | | Graph subsets | | | |
|---|---|---|---|---|---|
| | | Both constraints | | One constraint, one target | Both are targets |
| | | Same subset | Different subsets | | |
| | | ? $t_2, \lambda$:<br>$\boldsymbol{\beta}(\boldsymbol{f}(x_1, x_2) \mid x_1 \in \mathbf{G'},\ x_2 = t_2) \approx \mathbf{P}$;<br>$\mathbf{t_1}\ \lambda\ t_2$ | | pattern. This may involve only one lookup subtask[5] or two:<br><br>? $G'', t_2, \lambda, \psi$:<br>$\boldsymbol{\beta}(\boldsymbol{f}(x_1, x_2) \mid x_1 \in G'',\ x_2 = t_2) \approx \mathbf{P_2}$;<br>$(\mathbf{G'}, \mathbf{t_1})\ \lambda\ (G'', t_2)$;<br>$\mathbf{t_1} \psi\ t_2$<br><br>or<br><br>? $G'', t_1, \lambda, \psi$:<br>$\boldsymbol{\beta}(\boldsymbol{f}(x_1, x_2) \mid x_1 \in \mathbf{G'},\ x_2 = t_1) \approx \mathbf{P_1}$;<br>$\boldsymbol{\beta}(\boldsymbol{f}(x_1, x_2) \mid x_1 \in G'',\ x_2 = \mathbf{t_2}) \approx \mathbf{P_2}$;<br>$(\mathbf{G'}, t_1)\ \lambda\ (G'', t_2)$;<br>$t_1 \psi\ \mathbf{t_2}$ | ? $G', G'', t_2, \lambda, \psi$:<br>$\boldsymbol{\beta}(\boldsymbol{f}(x_1, x_2) \mid x_1 \in G',\ x_2 = \mathbf{t_1}) \approx \mathbf{P_1}$;<br>$\boldsymbol{\beta}(\boldsymbol{f}(x_1, x_2) \mid x_1 \in G'',\ x_2 = t_2) \approx \mathbf{P_2}$;<br>$(G', \mathbf{t_1})\ \lambda\ (G'', t_2)$;<br>$\mathbf{t_1} \psi\ t_2$ |
| | Both are targets | **Inverse comparison** of the time points at which two different patterns occur, over the same graph subset.<br><br>? $t_1, t_2, \lambda$:<br>$\boldsymbol{\beta}(\boldsymbol{f}(x_1, x_2) \mid x_1 \in \mathbf{G'},\ x_2 = t_1) \approx \mathbf{P_1}$;<br>$\boldsymbol{\beta}(\boldsymbol{f}(x_1, x_2) \mid x_1 \in \mathbf{G'},\ x_2 = t_2) \approx \mathbf{P_2}$;<br>$t_1\ \lambda\ t_2$ | **Inverse comparison** of the time points at which two different patterns occur, over two different graph subsets.<br><br>? $t_1, t_2, \lambda$:<br>$\boldsymbol{\beta}(\boldsymbol{f}(x_1, x_2) \mid x_1 \in \mathbf{G'},\ x_2 = t_1) \approx \mathbf{P_1}$;<br>$\boldsymbol{\beta}(\boldsymbol{f}(x_1, x_2) \mid x_1 \in \mathbf{G''},\ x_2 = t_2) \approx \mathbf{P_2}$;<br>$t_1\ \lambda\ t_2$ | **Inverse comparison** of the graph subsets associated with given patterns, where one of the graph subsets is specified, but the time at which it occurs is unknown, the other graph subset and time at which the pattern occurs is not specified. In addition, we may wish to compare the time points at which the patterns occurred.<br><br>? $G', G'', t_1, t_2, \lambda, \psi$:<br>$\boldsymbol{\beta}(\boldsymbol{f}(x_1, x_2) \mid x_1 \in \mathbf{G'},\ x_2 = t_1) \approx \mathbf{P_1}$;<br>$\boldsymbol{\beta}(\boldsymbol{f}(x_1, x_2) \mid x_1 \in G'',\ x_2 = t_2) \approx \mathbf{P_2}$;<br>$(\mathbf{G'}, t_1)\ \lambda\ (G'', t_2)$;<br>$t_1 \psi\ t_2$ | **Inverse comparison** of the graph subsets and time points associated with two given patterns.<br><br>? $G', G'', t_1, t_2, \lambda, \psi$:<br>$\boldsymbol{\beta}(\boldsymbol{f}(x_1, x_2) \mid x_1 \in G',\ x_2 = t_1) \approx \mathbf{P_1}$;<br>$\boldsymbol{\beta}(\boldsymbol{f}(x_1, x_2) \mid x_1 \in G'',\ x_2 = t_2) \approx \mathbf{P_2}$;<br>$(G', t_1)\ \lambda\ (G'', t_2)$;<br>$t_1 \psi\ t_2$ |

[5] The first task is reduced from:? $G'', t2, \lambda, \psi$: $\boldsymbol{\beta}(\boldsymbol{f}(x_1, x_2) \mid x_1 \in \mathbf{G'},\ x_2 = \mathbf{t_1}) \approx \mathbf{P_1}$; $\boldsymbol{\beta}(\boldsymbol{f}(x_1, x_2) \mid x_1 \in G'',\ x_2 = t_2) \approx \mathbf{P_2}$;$(\mathbf{G'}, \mathbf{t_1})\ \lambda\ (G'', t_2)$; $\mathbf{t_1} \psi\ t_2$ i.e. all information (the graph subset, time point and pattern) is known in the first lookup subtask.

**Figure 7 Comparison quadrant 3: considers comparisons involving the behaviour of an attribute of a single graph element over time (i.e. a temporal trend)**

| | | | Graph elements (nodes, edges, graph objects) | | | |
|---|---|---|---|---|---|---|
| | | | Both graph elements specified | | One graph element specified | Neither graph element specified |
| | | | Single/same graph element | Two different graph elements | | |
| Time intervals | Both Constraints | Same interval | **Direct comparison** of the attribute patterns *of two different attributes* of the same graph element over the same time interval.<br><br>?$p_1, p_2, \lambda$:<br>$\boldsymbol{\beta}(\boldsymbol{f_1}(x_1, x_2) \mid x_1 = \mathbf{g}, x_2 \in \mathbf{T'}) \approx p_1$;<br>$\boldsymbol{\beta}(\boldsymbol{f_2}(x_1, x_2) \mid x_1 = \mathbf{g}, x_2 \in \mathbf{T'}) \approx p_2$;<br>$p_1 \lambda\, p_2$ | **Direct comparison** of the patterns of two different graph elements over the same time interval<br><br>?$p_1, p_2, \lambda$:<br>$\boldsymbol{\beta}(\boldsymbol{f}(x_1, x_2) \mid x_1 = \mathbf{g_1}, x_2 \in \mathbf{T'}) \approx p_1$;<br>$\boldsymbol{\beta}(\boldsymbol{f}(x_1, x_2) \mid x_1 = \mathbf{g_2}, x_2 \in \mathbf{T'}) \approx p_2$;<br>$p_1 \lambda\, p_2$ | **Inverse comparison** of a graph element associated with a given pattern over a given time interval, with a given graph element.[6]<br><br>?$g_2, \lambda$:<br>$\boldsymbol{\beta}(\boldsymbol{f}(x_1, x_2) \mid x_1 = g_2, x_2 \in \mathbf{T'}) \approx \mathbf{P}$;<br>$\mathbf{g_1} \lambda\, g_2$ | **Inverse comparison** of two graph elements associated with given patterns over the same given time interval.<br><br>?$g_1, g_2, \lambda$:<br>$\boldsymbol{\beta}(\boldsymbol{f}(x_1, x_2) \mid x_1 = g_1, x_2 \in \mathbf{T'}) \approx \mathbf{P_1}$;<br>$\boldsymbol{\beta}(\boldsymbol{f}(x_1, x_2) \mid x_1 = g_2, x_2 \in \mathbf{T'}) \approx \mathbf{P_2}$;<br>$g_1 \lambda\, g_2$ |
| | | Different intervals | **Direct comparison** of the patterns of the same graph element over two different time intervals.<br><br>?$p_1, p_2, \lambda$:<br>$\boldsymbol{\beta}(\boldsymbol{f}(x_1, x_2) \mid x_1 = \mathbf{g}, x_2 \in \mathbf{T'}) \approx p_1$;<br>$\boldsymbol{\beta}(\boldsymbol{f}(x_1, x_2) \mid x_1 = \mathbf{g}, x_2 \in \mathbf{T''}) \approx p_2$;<br>$p_1 \lambda\, p_2$ | **Direct comparison** of the patterns of two different graph elements over two different time intervals.<br><br>?$p_1, p_2, \lambda$:<br>$\boldsymbol{\beta}(\boldsymbol{f}(x_1, x_2) \mid x_1 = \mathbf{g_1}, x_2 \in \mathbf{T'}) \approx p_1$;<br>$\boldsymbol{\beta}(\boldsymbol{f}(x_1, x_2) \mid x_1 = \mathbf{g_2}, x_2 \in \mathbf{T''}) \approx p_2$;<br>$p_1 \lambda\, p_2$ | **Inverse comparison** as above[7]. | **Inverse comparison** of two graph elements associated with given patterns over the two different given time intervals.<br><br>?$g_1, g_2, \lambda$:<br>$\boldsymbol{\beta}(\boldsymbol{f}(x_1, x_2) \mid x_1 = g_1, x_2 \in \mathbf{T'}) \approx \mathbf{P_1}$;<br>$\boldsymbol{\beta}(\boldsymbol{f}(x_1, x_2) \mid x_1 = g_2, x_2 \in \mathbf{T''}) \approx \mathbf{P_2}$;<br>$g_1 \lambda\, g_2$ |
| | One constraint, one target | | **Inverse comparison** of a time interval associated over which a given pattern occurs for a given graph element, with a specified time interval.[8]<br><br>?$T'', \lambda$: | **Inverse comparison** as left[9]. | **Inverse comparison** of a given graph element with a graph element associated with a given pattern (over a time interval which may or may not be specified) *and* comparison of a given time interval with a time interval associated with a given pattern (which | **Inverse comparison** of two graph elements associated with given patterns (one of which is a pattern over a specified time interval) *and* comparison of the time intervals over which the patterns occur.<br><br>?$g_1, g_2, T'', \lambda, \psi$: |

[6] Reduced from: ?$g_2, \lambda$: $\boldsymbol{\beta}(\boldsymbol{f}(x_1, x_2) \mid x_1 = \mathbf{g_1}, x_2 \in \mathbf{T'}) \approx \mathbf{P_1}$; $\boldsymbol{\beta}(\boldsymbol{f}(x_1, x_2) \mid x_1 = g_2, x_2 \in \mathbf{T'}) \approx \mathbf{P_2}$; $\mathbf{g_1} \lambda\, g_2$

[7] Reduced from ?$g_2, \lambda$: $\boldsymbol{\beta}(\boldsymbol{f}(x_1, x_2) \mid x_1 = \mathbf{g_1}, x_2 \in \mathbf{T'}) \approx \mathbf{P_1}$; $\boldsymbol{\beta}(\boldsymbol{f}(x_1, x_2) \mid x_1 = g_2, x_2 \in \mathbf{T''}) \approx \mathbf{P_2}$; $\mathbf{g_1} \lambda\, g_2$

[8] Reduced from: ?$T'', \lambda$: $\boldsymbol{\beta}(\boldsymbol{f}(x_1, x_2) \mid x_1 = \mathbf{g}, x_2 \in \mathbf{T'}) \approx \mathbf{P_1}$; $\boldsymbol{\beta}(\boldsymbol{f}(x_1, x_2) \mid x_1 = \mathbf{g}, x_2 \in T'') \approx \mathbf{P_2}$; $\mathbf{T'} \lambda\, T''$

[9] Reduced from: ?$T'', \lambda$: $\boldsymbol{\beta}(\boldsymbol{f}(x_1, x_2) \mid x_1 = \mathbf{g_1}, x_2 \in \mathbf{T'}) \approx \mathbf{P_1}$; $\boldsymbol{\beta}(\boldsymbol{f}(x_1, x_2) \mid x_1 = \mathbf{g_2}, x_2 \in T'') \approx \mathbf{P_2}$; $\mathbf{T'} \lambda\, T''$

| | Graph elements (nodes, edges, graph objects) | | | |
|---|---|---|---|---|
| | Both graph elements specified | | One graph element specified | Neither graph element specified |
| | Single/same graph element | Two different graph elements | | |
| | $\boldsymbol{\beta}(\boldsymbol{f}(x_1, x_2) \mid x_1 = \mathbf{g}, x_2 \in T'') \approx \mathbf{P}$;<br>$\mathbf{T'}\ \lambda\ T''$ | | may or may not be associated with a given graph element). This may involve only one lookup subtask[10] or two:<br><br>?$g_2, T'', \lambda, \psi$:<br>$\boldsymbol{\beta}(\boldsymbol{f}(x_1, x_2) \mid x_1 = g_2, x_2 \in T'') \approx \mathbf{P}$;<br>$\mathbf{g_1}\ \lambda\ g_2$;<br>$\mathbf{T'}\ \psi\ T''$<br><br>Or<br><br>?$g_2, T', \lambda, \psi$:<br>$\boldsymbol{\beta}(\boldsymbol{f}(x_1, x_2) \mid x_1 = \mathbf{g_1}, x_2 \in T') \approx \mathbf{P_1}$;<br>$\boldsymbol{\beta}(\boldsymbol{f}(x_1, x_2) \mid x_1 = g_2, x_2 \in \mathbf{T''}) \approx \mathbf{P_2}$;<br>$\mathbf{g_1}\ \lambda\ g_2$;<br>$T'\ \psi\ \mathbf{T''}$ | $\boldsymbol{\beta}(\boldsymbol{f}(x_1, x_2) \mid x_1 = g_1, x_2 \in \mathbf{T'}) \approx \mathbf{P_1}$;<br>$\boldsymbol{\beta}(\boldsymbol{f}(x_1, x_2) \mid x_1 = g_2, x_2 \in T'') \approx \mathbf{P_2}$;<br>$g_1\ \lambda\ g_2$;<br>$\mathbf{T'}\ \psi\ T''$ |
| Both are targets | **Inverse comparison** of the time intervals over which the given patterns occur for a single given graph element.<br><br>?$T', T'', \lambda$:<br>$\boldsymbol{\beta}(\boldsymbol{f}(x_1, x_2) \mid x_1 = \mathbf{g}, x_2 \in T') \approx \mathbf{P_1}$;<br>$\boldsymbol{\beta}(\boldsymbol{f}(x_1, x_2) \mid x_1 = \mathbf{g}, x_2 \in T'') \approx \mathbf{P_2}$;<br>$T'\ \lambda\ T''$ | **Inverse comparison** of the time intervals over which the given patterns occur for two different graph elements.<br><br>?$T', T'', \lambda$:<br>$\boldsymbol{\beta}(\boldsymbol{f}(x_1, x_2) \mid x_1 = \mathbf{g_1}, x_2 \in T') \approx \mathbf{P_1}$;<br>$\boldsymbol{\beta}(\boldsymbol{f}(x_1, x_2) \mid x_1 = \mathbf{g_2}, x_2 \in T'') \approx \mathbf{P_2}$;<br>$T'\ \lambda\ T''$ | **Inverse comparison** of a specified graph element and a graph element associated with a given pattern (over an unspecified time interval) *and* comparison of the time intervals over which the patterns occur.<br><br>? $g_2, T', T'', \lambda, \psi$:<br>$\boldsymbol{\beta}(\boldsymbol{f}(x_1, x_2) \mid x_1 = \mathbf{g_1}, x_2 \in T') \approx \mathbf{P_1}$;<br>$\boldsymbol{\beta}(\boldsymbol{f}(x_1, x_2) \mid x_1 = g_2, x_2 \in T'') \approx \mathbf{P_2}$;<br>$\mathbf{g_1}\ \lambda\ g_2$;<br>$T'\ \psi\ T''$ | **Inverse comparison** of graph elements and time intervals associated with two given patterns.<br><br>?$g_1, g_2, T', T'', \lambda, \psi$:<br>$\boldsymbol{\beta}(\boldsymbol{f}(x_1, x_2) \mid x_1 = g_1, x_2 \in T') \approx \mathbf{P_1}$;<br>$\boldsymbol{\beta}(\boldsymbol{f}(x_1, x_2) \mid x_1 = g_2, x_2 \in T'') \approx \mathbf{P_2}$;<br>$g_1\ \lambda\ g_2$;<br>$T'\ \psi\ T''$ |

[10] Reduced from: ?$g_2, T', \lambda, \psi$: $\boldsymbol{\beta}(\boldsymbol{f}(x_1, x_2) \mid x_1 = \mathbf{g_1}, x_2 \in \mathbf{T'}) \approx \mathbf{P_1}$; $\boldsymbol{\beta}(\boldsymbol{f}(x_1, x_2) \mid x_1 = g_2, x_2 \in T'') \approx \mathbf{P_2}$; $\mathbf{g_1}\ \lambda\ g_2$; $\mathbf{T'}\ \psi\ T''$

Figure 8 Comparison quadrant 4: considers comparisons involving aspectual behaviours (i) the behaviour of temporal trends for all graph elements, over the graph (ii) the behaviour of an attribute over the graph, over time

| | Graph subsets | | | |
|---|---|---|---|---|
| | Both graph subsets specified | | One graph subset specified | Neither graph subset specified |
| | Single/same subset | Two different subsets | | |
| Time intervals / Both constraints / Same time | **Direct comparison** of distributions of temporal trends over the graph *for two different attributes* over the same time interval and for the same graph subset:<br><br>? $p_1, p_2, \lambda$:<br>$\boldsymbol{\beta}_G\{\boldsymbol{\beta}_T[\boldsymbol{f_1}(x_1, x_2) \mid x_2 \in \mathbf{T'})] \mid x_1 \in \mathbf{G'}\} \approx p_1$;<br>$\boldsymbol{\beta}_G\{\boldsymbol{\beta}_T[\boldsymbol{f_2}(x_1, x_2) \mid x_2 \in \mathbf{T'})] \mid x_1 \in \mathbf{G'}\} \approx p_2$;<br>$p_1 \lambda\, p_2$<br><br>Or<br><br>temporal trends in distributions of an attribute over the graph *for two different attributes* for the same graph subset and over the same time interval:<br><br>? $p_1, p_2, \lambda$:<br>$\boldsymbol{\beta}_T\{\boldsymbol{\beta}_G[\boldsymbol{f_1}(x_1, x_2) \mid x_1 \in \mathbf{G'})] \mid x_2 \in \mathbf{T'}\} \approx p_1$;<br>$\boldsymbol{\beta}_T\{\boldsymbol{\beta}_G[\boldsymbol{f_2}(x_1, x_2) \mid x_1 \in \mathbf{G'})] \mid x_2 \in \mathbf{T'}\} \approx p_2$;<br>$p_1 \lambda\, p_2$ | **Direct comparison** of distributions of temporal trends over two different graph subsets over the same time interval:<br><br>? $p_1, p_2, \lambda$:<br>$\boldsymbol{\beta}_G\{\boldsymbol{\beta}_T[\boldsymbol{f_1}(x_1, x_2) \mid x_2 \in \mathbf{T'})] \mid x_1 \in \mathbf{G'}\} \approx p_1$;<br>$\boldsymbol{\beta}_G\{\boldsymbol{\beta}_T[\boldsymbol{f_2}(x_1, x_2) \mid x_2 \in \mathbf{T'})] \mid x_1 \in \mathbf{G''}\} \approx p_2$;<br>$p_1 \lambda\, p_2$<br><br>Or<br><br>temporal trends in distributions of an attribute over the graph, over two different graph subsets over the same time interval:<br><br>? $p_1, p_2, \lambda$:<br>$\boldsymbol{\beta}_T\{\boldsymbol{\beta}_G[\boldsymbol{f_1}(x_1, x_2) \mid x_1 \in \mathbf{G'})] \mid x_2 \in \mathbf{T'}\} \approx p_1$;<br>$\boldsymbol{\beta}_T\{\boldsymbol{\beta}_G[\boldsymbol{f_2}(x_1, x_2) \mid x_1 \in \mathbf{G''})] \mid x_2 \in \mathbf{T'}\} \approx p_2$;<br>$p_1 \lambda\, p_2$ | **Inverse comparison** of the subset of graph elements associated with a given pattern involving a given time interval, and a given subset of graph elements[11]:<br><br>? $G'', \lambda$:<br>$\boldsymbol{\beta}_G\{\boldsymbol{\beta}_T[\boldsymbol{f}(x_1, x_2) \mid x_2 \in \mathbf{T'})] \mid x_1 \in G''\} \approx \mathbf{P}$;<br>$\mathbf{G'}\ \lambda\ G''$;<br><br>or<br><br>?$G'', \lambda$:<br>$\boldsymbol{\beta}_T\{\boldsymbol{\beta}_G[\boldsymbol{f}(x_1, x_2) \mid x_1 \in G'')] \mid x_2 \in \mathbf{T'}\} \approx \mathbf{P}$;<br>$\mathbf{G'}\ \lambda\ G''$; | **Inverse comparison** of two graph subsets associated with two given patterns involving the same time interval:<br><br>? $G', G'', \lambda$:<br>$\boldsymbol{\beta}_G\{\boldsymbol{\beta}_T[\boldsymbol{f}(x_1, x_2) \mid x_2 \in \mathbf{T'})] \mid x_1 \in G'\} \approx \mathbf{P_1}$;<br>$\boldsymbol{\beta}_G\{\boldsymbol{\beta}_T[\boldsymbol{f}(x_1, x_2) \mid x_2 \in \mathbf{T'})] \mid x_1 \in G''\} \approx \mathbf{P_2}$;<br>$G'\ \lambda\ G''$;<br><br>or<br><br>? $G', G'', \lambda$:<br>$\boldsymbol{\beta}_T\{\boldsymbol{\beta}_G[\boldsymbol{f}(x_1, x_2) \mid x_1 \in G')] \mid x_2 \in \mathbf{T'}\} \approx \mathbf{P_1}$;<br>$\boldsymbol{\beta}_T\{\boldsymbol{\beta}_G[\boldsymbol{f}(x_1, x_2) \mid x_1 \in G'')] \mid x_2 \in \mathbf{T'}\} \approx \mathbf{P_2}$;<br>$G'\ \lambda\ G''$; |

[11] Reduced from: ? $G'', \lambda$: $\boldsymbol{\beta}_G\{\boldsymbol{\beta}_T[\boldsymbol{f}(x_1, x_2) \mid x_2 \in \mathbf{T'})] \mid x_1 \in \mathbf{G'}\} \approx \mathbf{P_1}$; $\boldsymbol{\beta}_G\{\boldsymbol{\beta}_T[\boldsymbol{f}(x_1, x_2) \mid x_2 \in \mathbf{T'})] \mid x_1 \in G''\} \approx \mathbf{P_2}$;$\mathbf{G'}\ \lambda\ G''$;
OR
?$G'', \lambda$: $\boldsymbol{\beta}_T\{\boldsymbol{\beta}_G[\boldsymbol{f}(x_1, x_2) \mid x_1 \in \mathbf{G'})] \mid x_2 \in \mathbf{T'}\} \approx \mathbf{P_1}$;$\boldsymbol{\beta}_T\{\boldsymbol{\beta}_G[\boldsymbol{f}(x_1, x_2) \mid x_1 \in G'')] \mid x_2 \in \mathbf{T'}\} \approx \mathbf{P_2}$; $\mathbf{G'}\ \lambda\ G''$;

| | Graph subsets | | | |
|---|---|---|---|---|
| | Both graph subsets specified | | One graph subset specified | Neither graph subset specified |
| | Single/same subset | Two different subsets | | |
| Different times | **Direct comparison** of distributions of temporal trends over the graph for the same graph subset during two different time intervals:<br><br>? $p_1, p_2, \lambda$:<br>$\boldsymbol{\beta}_G\{\boldsymbol{\beta}_T[\boldsymbol{f}_1(x_1, x_2) \mid x_2 \in \mathbf{T'})] \mid x_1 \in \mathbf{G'}\} \approx p_1$;<br>$\boldsymbol{\beta}_G\{\boldsymbol{\beta}_T[\boldsymbol{f}_2(x_1, x_2) \mid x_2 \in \mathbf{T''})] \mid x_1 \in \mathbf{G'}\} \approx p_2$;<br>$p_1 \lambda\, p_2$<br><br>Or<br><br>temporal trends in distributions of an attribute over the graph, for the same graph subset over two different time intervals:<br><br>? $p_1, p_2, \lambda$:<br>$\boldsymbol{\beta}_T\{\boldsymbol{\beta}_G[\boldsymbol{f}_1(x_1, x_2) \mid x_1 \in \mathbf{G'})] \mid x_2 \in \mathbf{T'}\} \approx p_1$;<br>$\boldsymbol{\beta}_T\{\boldsymbol{\beta}_G[\boldsymbol{f}_2(x_1, x_2) \mid x_1 \in \mathbf{G'})] \mid x_2 \in \mathbf{T''}\} \approx p_2$;<br>$p_1 \lambda\, p_2$ | **Direct comparison** of distributions of temporal trends over two different graph subsets over two different time intervals:<br><br>? $p_1, p_2, \lambda$:<br>$\boldsymbol{\beta}_G\{\boldsymbol{\beta}_T[\boldsymbol{f}_1(x_1, x_2) \mid x_2 \in \mathbf{T'})] \mid x_1 \in \mathbf{G'}\} \approx p_1$;<br>$\boldsymbol{\beta}_G\{\boldsymbol{\beta}_T[\boldsymbol{f}_2(x_1, x_2) \mid x_2 \in \mathbf{T''})] \mid x_1 \in \mathbf{G''}\} \approx p_2$;<br>$p_1 \lambda\, p_2$<br><br>Or<br><br>temporal trends in distributions of an attribute over the graph, over two different graph subsets over two different time intervals:<br><br>? $p_1, p_2, \lambda$:<br>$\boldsymbol{\beta}_T\{\boldsymbol{\beta}_G[\boldsymbol{f}_1(x_1, x_2) \mid x_1 \in \mathbf{G'})] \mid x_2 \in \mathbf{T'}\} \approx p_1$;<br>$\boldsymbol{\beta}_T\{\boldsymbol{\beta}_G[\boldsymbol{f}_2(x_1, x_2) \mid x_1 \in \mathbf{G''})] \mid x_2 \in \mathbf{T''}\} \approx p_2$;<br>$p_1 \lambda\, p_2$ | **Inverse comparison** as above[12] | **Inverse comparison** of two graph subsets associated with two given patterns involving two different time intervals<br><br>? $G', G'', \lambda$:<br>$\boldsymbol{\beta}_G\{\boldsymbol{\beta}_T[\boldsymbol{f}(x_1, x_2) \mid x_2 \in \mathbf{T'})] \mid x_1 \in G'\} \approx \mathbf{P}_1$;<br>$\boldsymbol{\beta}_G\{\boldsymbol{\beta}_T[\boldsymbol{f}(x_1, x_2) \mid x_2 \in \mathbf{T''})] \mid x_1 \in G''\} \approx \mathbf{P}_2$;<br>$G' \lambda\, G''$;<br><br>or<br><br>? $G', G'', \lambda$:<br>$\boldsymbol{\beta}_T\{\boldsymbol{\beta}_G[\boldsymbol{f}(x_1, x_2) \mid x_1 \in G')] \mid x_2 \in \mathbf{T'}\} \approx \mathbf{P}_1$;<br>$\boldsymbol{\beta}_T\{\boldsymbol{\beta}_G[\boldsymbol{f}(x_1, x_2) \mid x_1 \in G'')] \mid x_2 \in \mathbf{T''}\} \approx \mathbf{P}_2$;<br>$G' \lambda\, G''$; |

[12] Reduced from: ? $G'', \lambda$: $\boldsymbol{\beta}_G\{\boldsymbol{\beta}_T[\boldsymbol{f}(x_1, x_2) \mid x_2 \in \mathbf{T'})] \mid x_1 \in \mathbf{G'}\} \approx \mathbf{P}_1$; $\boldsymbol{\beta}_G\{\boldsymbol{\beta}_T[\boldsymbol{f}(x_1, x_2) \mid x_2 \in \mathbf{T''})] \mid x_1 \in G''\} \approx \mathbf{P}_2$; $\mathbf{G'}\, \lambda\, G''$;OR
? $G'', \lambda$: $\boldsymbol{\beta}_T\{\boldsymbol{\beta}_G[\boldsymbol{f}(x_1, x_2) \mid x_1 \in \mathbf{G'})] \mid x_2 \in \mathbf{T'}\} \approx \mathbf{P}_1$;$\boldsymbol{\beta}_T\{\boldsymbol{\beta}_G[\boldsymbol{f}(x_1, x_2) \mid x_1 \in G'')] \mid x_2 \in \mathbf{T''}\} \approx \mathbf{P}_2$; $\mathbf{G'}\, \lambda\, G''$;

<table>
<tr><th rowspan="3"></th><th colspan="4">Graph subsets</th></tr>
<tr><th colspan="2">Both graph subsets specified</th><th rowspan="2">One graph subset specified</th><th rowspan="2">Neither graph subset specified</th></tr>
<tr><th>Single/same subset</th><th>Two different subsets</th></tr>
<tr>
<td>One constraint, one target</td>
<td>Inverse comparison of a time interval associated with a given pattern and graph subset, and a given time interval[13]:<br><br>? $T''$, $\lambda$:<br>$\boldsymbol{\theta}_G\{\boldsymbol{\theta}_T[\boldsymbol{f}(x_1, x_2) \mid x_2 \in T'')] \mid x_1 \in \mathbf{G'}\} \approx \mathbf{P}$;<br>$\mathbf{T'}\ \lambda\ T''$;<br><br>or<br><br>? $T''$, $\lambda$:<br>$\boldsymbol{\theta}_T\{\boldsymbol{\theta}_G[\boldsymbol{f}(x_1, x_2) \mid x_1 \in \mathbf{G'})] \mid x_2 \in T''\} \approx \mathbf{P}$;<br>$\mathbf{T'}\ \lambda\ T''$;</td>
<td>Inverse comparison as left[14].</td>
<td>Inverse comparison of a time interval and graph subset associated with a given pattern, with a given time interval and graph subset[15]<br><br>? $G''$, $T''$, $\lambda$, $\psi$:<br>$\boldsymbol{\theta}_G\{\boldsymbol{\theta}_T[\boldsymbol{f}(x_1, x_2) \mid x_2 \in T'')] \mid x_1 \in G''\} \approx \mathbf{P}$;<br>$\mathbf{T'}\ \lambda\ T''$;<br>$\mathbf{G'}\ \psi\ G''$;<br><br>or<br><br>? $G''$, $T''$, $\lambda$, $\psi$:<br>$\boldsymbol{\theta}_T\{\boldsymbol{\theta}_G[\boldsymbol{f}(x_1, x_2) \mid x_1 \in G'')] \mid x_2 \in T''\} \approx \mathbf{P}$;<br>$\mathbf{T'}\ \lambda\ T''$;<br>$\mathbf{G'}\ \psi\ G''$;<br><br>OR<br><br>Inverse comparison of a graph object associated with a pattern involving a given time interval, and a given graph object and a time interval associated with a pattern</td>
<td>Inverse comparison of graph subsets and time intervals associated with two given patterns, where one of the patterns involves a given time interval:<br><br>? $G'$, $G''$, $T''$, $\lambda$, $\psi$:<br>$\boldsymbol{\theta}_G\{\boldsymbol{\theta}_T[\boldsymbol{f}(x_1, x_2) \mid x_2 \in \mathbf{T'})] \mid x_1 \in G'\} \approx \mathbf{P_1}$;<br>$\boldsymbol{\theta}_G\{\boldsymbol{\theta}_T[\boldsymbol{f}(x_1, x_2) \mid x_2 \in T'')] \mid x_1 \in G''\} \approx \mathbf{P_2}$;<br>$\mathbf{T'}\ \lambda\ T''$;<br>$G'\ \psi\ G''$;<br><br>or<br><br>? $G'$, $G''$, $T''$, $\lambda$, $\psi$:<br>$\boldsymbol{\theta}_T\{\boldsymbol{\theta}_G[\boldsymbol{f}(x_1, x_2) \mid x_1 \in G')] \mid x_2 \in \mathbf{T'}\} \approx \mathbf{P_1}$;<br>$\boldsymbol{\theta}_T\{\boldsymbol{\theta}_G[\boldsymbol{f}(x_1, x_2) \mid x_1 \in G'')] \mid x_2 \in T''\} \approx \mathbf{P_2}$;<br>$\mathbf{T'}\ \lambda\ T''$;<br>$G'\ \psi\ G''$;</td>
</tr>
</table>

[13] Reduced from: ? $T''$, $\lambda$: $\boldsymbol{\theta}_G\{\boldsymbol{\theta}_T[\boldsymbol{f}(x_1, x_2) \mid x_2 \in \mathbf{T'})] \mid x_1 \in \mathbf{G'}\} \approx \mathbf{P_1}$; $\boldsymbol{\theta}_G\{\boldsymbol{\theta}_T[\boldsymbol{f}(x_1, x_2) \mid x_2 \in T'')] \mid x_1 \in \mathbf{G'}\} \approx \mathbf{P_2}$; $\mathbf{T'}\ \lambda\ T''$; or ? $T''$, $\lambda$: $\boldsymbol{\theta}_T\{\boldsymbol{\theta}_G[\boldsymbol{f}(x_1, x_2) \mid x_1 \in \mathbf{G'})] \mid x_2 \in \mathbf{T'}\} \approx \mathbf{P_1}$; $\boldsymbol{\theta}_T\{\boldsymbol{\theta}_G[\boldsymbol{f}(x_1, x_2) \mid x_1 \in \mathbf{G'})] \mid x_2 \in T''\} \approx \mathbf{P_2}$;
$\mathbf{T'}\ \lambda\ T''$;

[14] Reduced from: ? $T''$, $\lambda$: $\boldsymbol{\theta}_G\{\boldsymbol{\theta}_T[\boldsymbol{f}(x_1, x_2) \mid x_2 \in \mathbf{T'})] \mid x_1 \in \mathbf{G'}\} \approx \mathbf{P_1}$; $\boldsymbol{\theta}_G\{\boldsymbol{\theta}_T[\boldsymbol{f}(x_1, x_2) \mid x_2 \in T'')] \mid x_1 \in \mathbf{G''}\} \approx \mathbf{P_2}$;$\mathbf{T'}\ \lambda\ T''$; or ? $T''$, $\lambda$: $\boldsymbol{\theta}_T\{\boldsymbol{\theta}_G[\boldsymbol{f}(x_1, x_2) \mid x_1 \in \mathbf{G'})] \mid x_2 \in \mathbf{T'}\} \approx \mathbf{P_1}$;$\boldsymbol{\theta}_T\{\boldsymbol{\theta}_G[\boldsymbol{f}(x_1, x_2) \mid x_1 \in \mathbf{G''})] \mid x_2 \in T''\} \approx \mathbf{P_2}$;
$\mathbf{T'}\ \lambda\ T''$;

[15] Reduced from: ? $G''$, $T''$, $\lambda$, $\psi$: $\boldsymbol{\theta}_G\{\boldsymbol{\theta}_T[\boldsymbol{f}(x_1, x_2) \mid x_2 \in \mathbf{T'})] \mid x_1 \in \mathbf{G'}\} \approx \mathbf{P_1}$; $\boldsymbol{\theta}_G\{\boldsymbol{\theta}_T[\boldsymbol{f}(x_1, x_2) \mid x_2 \in T'')] \mid x_1 \in G''\} \approx \mathbf{P_2}$; $\mathbf{T'}\ \lambda\ T''$; $\mathbf{G'}\ \psi\ G''$; or ? $G''$, $T''$, $\lambda$, $\psi$: $\boldsymbol{\theta}_T\{\boldsymbol{\theta}_G[\boldsymbol{f}(x_1, x_2) \mid x_1 \in \mathbf{G'})] \mid x_2 \in \mathbf{T'}\} \approx \mathbf{P_1}$; $\boldsymbol{\theta}_T\{\boldsymbol{\theta}_G[\boldsymbol{f}(x_1, x_2) \mid x_1 \in G'')] \mid x_2 \in T''\} \approx \mathbf{P_2}$; $\mathbf{T'}\ \lambda\ T''$; $\mathbf{G'}\ \psi\ G''$;

| | | Graph subsets | | | |
|---|---|---|---|---|---|
| | | Both graph subsets specified | | One graph subset specified | Neither graph subset specified |
| | | Single/same subset | Two different subsets | | |
| | | | | involving a given graph subset, and a given time interval.<br><br>? *G′, T′, T″, λ, ψ*:<br>$\beta_G\{\beta_T[f(x_1, x_2) \mid x_2 \in \mathbf{T'})] \mid x_1 \in G'\} \approx \mathbf{P_1}$;<br>$\beta_G\{\beta_T[f(x_1, x_2) \mid x_2 \in T'')] \mid x_1 \in \mathbf{G''}\} \approx \mathbf{P_2}$;<br>$\mathbf{T'}\ \lambda\ T''$;<br>$G'\ \psi\ \mathbf{G''}$;<br><br>or<br><br>? *G″, T′, λ, ψ*:<br>$\beta_T\{\beta_G[f(x_1, x_2) \mid x_1 \in \mathbf{G'})] \mid x_2 \in T'\} \approx \mathbf{P_1}$;<br>$\beta_T\{\beta_G[f(x_1, x_2) \mid x_1 \in G'')] \mid x_2 \in \mathbf{T''}\} \approx \mathbf{P_2}$;<br>$T'\ \lambda\ \mathbf{T''}$;<br>$\mathbf{G'}\ \psi\ G''$; | |
| | Both are targets | **Inverse comparison** of two time intervals associated with two given patterns involving the same graph subset:<br><br>? *T′, T″, λ*:<br>$\beta_G\{\beta_T[f(x_1, x_2) \mid x_2 \in T')] \mid x_1 \in \mathbf{G'}\} \approx \mathbf{P_1}$;<br>$\beta_G\{\beta_T[f(x_1, x_2) \mid x_2 \in T'')] \mid x_1 \in \mathbf{G'}\} \approx \mathbf{P_2}$;<br>$T'\ \lambda\ T''$;<br><br>or<br><br>? *T′, T″, λ*:<br>$\beta_T\{\beta_G[f(x_1, x_2) \mid x_1 \in \mathbf{G'})] \mid x_2 \in T'\} \approx \mathbf{P_1}$;<br>$\beta_T\{\beta_G[f(x_1, x_2) \mid x_1 \in \mathbf{G'})] \mid x_2 \in T''\} \approx \mathbf{P_2}$;<br>$T'\ \lambda\ T''$; | **Inverse comparison** of two time intervals associated with two given patterns involving two different graph subsets:<br><br>? *T′, T″, λ*:<br>$\beta_G\{\beta_T[f(x_1, x_2) \mid x_2 \in T')] \mid x_1 \in \mathbf{G'}\} \approx \mathbf{P_1}$;<br>$\beta_G\{\beta_T[f(x_1, x_2) \mid x_2 \in T'')] \mid x_1 \in \mathbf{G''}\} \approx \mathbf{P_2}$;<br>$T'\ \lambda\ T''$;<br><br>or<br><br>? *T′, T″, λ*:<br>$\beta_T\{\beta_G[f(x_1, x_2) \mid x_1 \in \mathbf{G'})] \mid x_2 \in T'\} \approx \mathbf{P_1}$;<br>$\beta_T\{\beta_G[f(x_1, x_2) \mid x_1 \in \mathbf{G''})] \mid x_2 \in T''\} \approx \mathbf{P_2}$; | **Inverse comparison** of graph subsets and time intervals associated with given patterns, where one of the patterns involves a given graph subset:<br><br>? *G″, T′, T″, λ, ψ*:<br>$\beta_G\{\beta_T[f(x_1, x_2) \mid x_2 \in T')] \mid x_1 \in \mathbf{G'}\} \approx \mathbf{P_1}$;<br>$\beta_G\{\beta_T[f(x_1, x_2) \mid x_2 \in T'')] \mid x_1 \in G''\} \approx \mathbf{P_2}$;<br>$T'\ \lambda\ T''$;<br>$\mathbf{G'}\ \psi\ G''$;<br><br>or<br><br>?*G″, T′, T″, λ, ψ:*<br>$\beta_T\{\beta_G[f(x_1, x_2) \mid x_1 \in \mathbf{G'})] \mid x_2 \in T'\} \approx \mathbf{P_1}$; | **Inverse comparison** of graph subsets and time intervals associated with given patterns:<br><br>? *G′, G″, T′, T″, λ, ψ*:<br>$\beta_G\{\beta_T[f(x_1, x_2) \mid x_2 \in T')] \mid x_1 \in G'\} \approx \mathbf{P_1}$;<br>$\beta_G\{\beta_T[f(x_1, x_2) \mid x_2 \in T'')] \mid x_1 \in G''\} \approx \mathbf{P_2}$;<br>$T'\ \lambda\ T''$;<br>$G'\ \psi\ G''$;<br><br>or<br><br>? *G′, G″, T′, T″, λ, ψ*:<br>$\beta_T\{\beta_G[f(x_1, x_2) \mid x_1 \in G')] \mid x_2 \in T'\} \approx \mathbf{P_1}$;<br>$\beta_T\{\beta_G[f(x_1, x_2) \mid x_1 \in G'')] \mid x_2 \in T''\} \approx \mathbf{P_2}$; |

| | | Graph subsets | | | |
|---|---|---|---|---|---|
| | | Both graph subsets specified | | One graph subset specified | Neither graph subset specified |
| | | Single/same subset | Two different subsets | | |
| | | | $T' \lambda T''$; | $\boldsymbol{\beta}_T\{\boldsymbol{\beta}_G[\boldsymbol{f}(x_1, x_2) \mid x_1 \in G'')] \mid x_2 \in T''\} \approx \mathbf{P_2}$;<br>$T' \lambda T''$;<br>$\mathbf{G'} \psi G''$; | $T' \lambda T''$;<br>$G' \psi G''$; |

### 4.1.3 Relation seeking

In relation seeking tasks, a relation between elements is given and the task is to find the elements related in the specified manner. In elementary relation seeking this generally involves finding attribute values related in a specified way, but may also involve a specified relation on time points and/or graph elements. Similarly in synoptic tasks, this involves finding patterns related in a specified way, but may also involve a specified relation on time intervals and/or graph subsets. The Andrienko framework makes an additional distinction between tasks involving the same or different attributes in the subtasks. The tasks in the matrices have been formulated to show the same attribute, but each task could also be formulated for the case where two different attributes are involved.

Note also that we do not show in the matrix tasks where attribute values or patterns are specified. These tasks can be formulated to produce tasks where either:

i. Both attribute values or patterns are specified. In this case, the relation seeking task will involve a specified relation on time points/intervals and/or graph elements/subsets. Taking an example from quadrant 2, we could have:

   Find graph subsets and time points associated with given patterns, where the graph subsets/time points are related in the specified way.

   $? \; G', G'', t_1, t_2:$
   $\boldsymbol{\beta}(\boldsymbol{f}(x_1, x_2) \mid x_1 \in G', x_2 = t_1) \approx \mathbf{P_1};$
   $\boldsymbol{\beta}(\boldsymbol{f}(x_1, x_2) \mid x_1 \in G'', x_2 = t_2) \approx \mathbf{P_2};$
   $t_1 \; \boldsymbol{\Psi} \; t_2;$
   $G' \; \boldsymbol{\Phi} \; G'';$

ii. One attribute value or pattern is specified. In this case, the specified relation may be between attribute values or patterns, graph elements or subsets and/or time points or intervals (as appropriate to the specified/unspecified elements in the task). Again, we give an example from quadrant 2:

   Find patterns related to a given pattern in the given way. Find also the graph subsets and time points over/at which the related patterns occur. A relation between graph subsets and/or time points may also be specified.

   $? \; G', G'', t_1, t_2, P_2:$
   $\boldsymbol{\beta}(\boldsymbol{f}(x_1, x_2) \mid x_1 \in G', x_2 = t_1) \approx \mathbf{P_1};$
   $\boldsymbol{\beta}(\boldsymbol{f}(x_1, x_2) \mid x_1 \in G'', x_2 = t_2) \approx P_2;$
   $t_1 \; \boldsymbol{\Psi} \; t_2;$
   $G' \; \boldsymbol{\Phi} \; G'';$
   $P_1 \; \boldsymbol{\Lambda} \; P_2$

Note that in the case where one of the subtasks is completely specified, we reduce the task to relation seeking involving a specified pattern or graph subset e.g.

$? \; G'', t_2, P_2:$
$\boldsymbol{\beta}(\boldsymbol{f}(x_1, x_2) \mid x_1 \in \mathbf{G'}, x_2 = \mathbf{t_1}) \approx \mathbf{P_1};$
$\boldsymbol{\beta}(\boldsymbol{f}(x_1, x_2) \mid x_1 \in G'', x_2 = t_2) \approx P_2;$

$\mathbf{t_1}\ \boldsymbol{\Psi}\ t_2;$
$\mathbf{G'}\ \boldsymbol{\Phi}\ G'';$
$\mathbf{P_1}\ \boldsymbol{\Lambda}\ P_2$

Can be reduced to:

> Find patterns/graph elements related in the given way to given patterns/graph elements. A relation on time points may also be specified.
> $?\ G'', t_2, P_2:$
> $\boldsymbol{\beta}(\boldsymbol{f}(x_1, x_2) \mid x_1 \in G'', x_2 = t_2) \approx P_2;$
> $\mathbf{t_1}\ \boldsymbol{\Psi}\ t_2;$
> $\mathbf{G'}\ \boldsymbol{\Phi}\ G'';$
> $\mathbf{P_1}\ \boldsymbol{\Lambda}\ P_2$

Again, we provide an overview of the relation seeking tasks based on the referential components involved. The permutations of tasks involving the same/different/specified/unspecified elements are given in the full task matrix.

**Figure 9 Relation seeking quadrant-level overview**

| | **Graph elements (nodes, edges, graph objects)** | **Graph subsets** |
|---|---|---|
| **Time points** | **Elementary**<br>$?\ t_1, t_2, g_1, g_2, y_1, y_2$:<br>$\boldsymbol{f}(t_1, g_1) = y_1;$<br>$\boldsymbol{f}(t_2, g_2) = y_2;$<br>$t_1\ \boldsymbol{\Psi}\ t_2;$<br>$g_1\ \boldsymbol{\Phi}\ g_2;$<br>$y_1\ \boldsymbol{\Lambda}\ y_2$<br><br>**Relation seeking** – find the attribute values related in the given manner (and possibly the corresponding graph element(s)/time point(s)). In this case the possible relation specified is domain dependent. Variations of this task depend on the number of time points and graph elements specified in the lookup sub tasks.<br><br>Additional constraints on the relations between graph elements and/or time points may also be specified. Depending on the elements involved in the lookup tasks (i.e. whether they are specified/unspecified, same or different), constraints may be any of the relations noted in the comparison matrix e.g.:<br><br>• between time points: equality (same/different time point), that time points are consecutive, occur before/after a given time point, that a certain distance exists between them etc.<br>• between graph elements: *equality (same/different element); set relations (between the sets of elements belonging to graph objects); equality of configuration (in graph objects); association (*where a single time point is specified in the lookup task or a constraint of equality is added on unspecified time points*)*. | **Synoptic**<br>$?\ G', G'', t_1, t_2, P_1, P_2:$<br>$\boldsymbol{\beta}(\boldsymbol{f}(x_1, x_2) \mid x_1 \in G',\ x_2 = t_1) \approx P_1;$<br>$\boldsymbol{\beta}(\boldsymbol{f}(x_1, x_2) \mid x_1 \in G'',\ x_2 = t_2) \approx P_2;$<br>$t_1\ \boldsymbol{\Psi}\ t_2;$<br>$G'\ \boldsymbol{\Phi}\ G'';$<br>$P_1\ \boldsymbol{\Lambda}\ P_2$<br><br>**Relation seeking –** find patterns of attribute(s) over the graph which are related in the given manner (and possibly the time point(s)/subsets of graph elements at/over which they occur). Possible specified relations between patterns are same (similar)/different/opposite. Variations depend on the number of time points and graph subsets specified in the lookup subtasks.<br><br>Additional constraints on relations between time points /or graph subsets may also be included in the task specification, depending on the elements involved in the lookup tasks. These are similar to the relations noted in the comparison matrix e.g.<br><br>• between time points: equality (same/different time point), that time points are consecutive, occur before/after a given time point, that a certain distance exists between them etc.<br>• between two graph subsets: equality (same/different subset); set relations (between the sets of nodes/edges belonging to the subset); equality of configuration of the subset, association (between nodes/graph objects, at a single time point only). |
| **Time intervals** | **Synoptic**<br>$?\ g_1, g_2; T', T'', P_1, P_2$:<br>$\boldsymbol{\beta}(\boldsymbol{f}(x_1, x_2) \mid x_1 = g_1, x_2 \in T') \approx P_1;$<br>$\boldsymbol{\beta}(\boldsymbol{f}(x_1, x_2) \mid x_1 = g_2, x_2 \in T'') \approx P_2;$<br>$g_1\ \boldsymbol{\Phi}\ g_2;\ T\ \boldsymbol{\Psi}\ T'';\ P_1\ \boldsymbol{\Lambda}\ P_2$ | **Synoptic**<br>$?\ G', G'', T', T'', P_1, P_2$:<br>$\boldsymbol{\beta}_G\{\boldsymbol{\beta}_T[\boldsymbol{f}(x_1, x_2) \mid x_2 \in T')] \mid x_1 \in G'\} \approx P_1;$<br>$\boldsymbol{\beta}_G\{\boldsymbol{\beta}_T[\boldsymbol{f}(x_1, x_2) \mid x_2 \in T'')] \mid x_1 \in G''\} \approx P_2;$<br>$T'\ \boldsymbol{\Psi}\ T'';\ G'\ \boldsymbol{\Phi}\ G'';\ P_1\ \boldsymbol{\Lambda}\ P_2$ |
| | **Relation seeking** – find the patterns of attribute(s) over time which are related in the given manner (and possibly find the graph element(s) to which they correspond/the time period(s) over which they occur). The possible specified relations between patterns are the same (similar)/different/opposite. Variations depend on the number of graph elements and time intervals specified in the lookup subtasks.<br><br>Additional constraints on relations between graph elements and/or time intervals may also be included in the task specification, depending on the elements involved in the lookup tasks. These are similar to the relations noted in the comparison matrix e.g.<br>• between graph elements: equality (same/different element); set relations (between the sets of elements belonging to graph objects); equality of configuration (in graph objects).<br>• Between the time intervals (over which the pattern occurs): happens before(/after), happens at the same time; between two intervals, or an instant and an interval: happens before(/after), starts, finishes, happens during; between intervals only: overlaps, meets [4]. | or<br><br>? $G', G'', T', T'', P_1, P_2$:<br>$\boldsymbol{\beta}_T\{\boldsymbol{\beta}_G[\boldsymbol{f}(x_1, x_2) \mid x_1 \in G')] \mid x_2 \in T'\} \approx P_1$;<br>$\boldsymbol{\beta}_T\{\boldsymbol{\beta}_G[\boldsymbol{f}(x_1, x_2) \mid x_1 \in G'')] \mid x_2 \in T''\} \approx P_2$;<br>$T' \boldsymbol{\Psi} T''; G' \boldsymbol{\Phi} G''; P_1 \boldsymbol{\Lambda} P_2$<br><br>**Relation seeking** – find (sub)patterns of either of the aspectual behaviours which are related in the given manner (and possibly find the graph subset/time interval associated with the found patterns). The possible specified relations between patterns are the same (similar)/different/opposite. Variations depend on the number of graph subsets and time intervals specified in the lookup subtasks.<br><br>Additional constraints on relations between time points /or graph subsets may also be included in the task specification, depending on the elements involved in the lookup tasks. These are similar to the relations noted in the comparison matrix e.g.<br>• between two graph subsets: equality (same/different subset); set relations (between the sets of nodes/edges belonging to the subset); equality of configuration of the subset, association (between nodes/graph objects, at a single time point only).<br>• Between the time intervals (over which the pattern occurs): happens before(/after), happens at the same time; between two intervals, or an instant and an interval: happens before(/after), starts, finishes, happens during; between intervals only: overlaps, meets [4]. |

**Figure 10 Relation seeking, quadrant 1: considers elementary relation seeking involving graph elements (nodes, edges, graph objects) and time points (i.e. the elementary comparison tasks)**

| | | | Graph elements (nodes, edges, graph objects) | | | |
|---|---|---|---|---|---|---|
| | | | **Both constraints** | | **One constraint, one target** | **Both are targets** |
| | | | **Same element** | **Different elements** | | |
| **Time points** | **Both constraints** | **Same time** | Not applicable | | Find the attribute value (and associated node) at a given time, which is related in the given way to an attribute value associated with a given graph object at the same given time point. A relation between graph elements may also be specified.<br><br>? $g_2, y_1, y_2$:<br>$\boldsymbol{f}(\mathbf{t}, \mathbf{g_1}) = y_1; \boldsymbol{f}(\mathbf{t}, g_2) = y_2;$<br>$y_1 \, \boldsymbol{\Lambda} \, y_2; \, \boldsymbol{g_1} \, \boldsymbol{\Phi} \, g_2$ | Find attribute values (and the nodes associated with them) at the same given time, which are related in the given way. A relation between graph elements may also be specified.<br><br>? $g_1, g_2, y_1, y_2$:<br>$\boldsymbol{f}(\mathbf{t}, g_1) = y_1; \boldsymbol{f}(\mathbf{t}, g_2) = y_2;$<br>$y_1 \, \boldsymbol{\Lambda} \, y_2; \, \boldsymbol{g_1} \, \boldsymbol{\Phi} \, g_2$ |
| | | **Different time** | | | Find the attribute value (and associated node) at a given time, which is related in the given way to an attribute value associated with a given graph object at a different given time point.<br><br>? $g_2, y_1, y_2$:<br>$\boldsymbol{f}(\mathbf{t_1}, \mathbf{g_1}) = y_1; \boldsymbol{f}(\mathbf{t_2}, g_2) = y_2;$<br>$y_1 \, \boldsymbol{\Lambda} \, y_2;$ | Find attribute values (and the nodes associated with them) at two given times, which are related in the given way. A relation between graph elements may also be specified.<br><br>? $g_1, g_2, y_1, y_2$:<br>$\boldsymbol{f}(\mathbf{t_1}, g_1) = y_1; \boldsymbol{f}(\mathbf{t_2}, g_2) = y_2;$<br>$y_1 \, \boldsymbol{\Lambda} \, y_2; \, \boldsymbol{g_1} \, \boldsymbol{\Phi} \, g_2$ |

| | Graph elements (nodes, edges, graph objects) | | | |
|---|---|---|---|---|
| | Both constraints | | One constraint, one target | Both are targets |
| | Same element | Different elements | | |
| One constraint, one target | Find an attribute value (and the time point at which it occurs) associated with a given graph element, which is related in the given way to an attribute value associated with a the same graph element at a given time. A relation between time points may also be specified.<br><br>? $t_2, y_1, y_2$:<br>$\boldsymbol{f}(\mathbf{t_1}, \mathbf{g}) = y_1; \boldsymbol{f}(t_2, \mathbf{g}) = y_2$;<br>$y_1 \boldsymbol{\Lambda} y_2$ | Find an attribute value (and the time point at which it occurs) associated with a given graph element, which is related in the given way to an attribute value associated with a different given graph element at a given time. A relation between time points may also be specified.<br><br>? $t_2, y_1, y_2$:<br>$\boldsymbol{f}(\mathbf{t_1}, \mathbf{g_1}) = y_1; \boldsymbol{f}(t_2, \mathbf{g_2}) = y_2$;<br>$t_1 \boldsymbol{\Psi} t_2$; $y_1 \boldsymbol{\Lambda} y$ | Find an attribute value (and the time point and graph element for which it occurs) related in the given way to an attribute value which is associated with a given graph element at a given time point. Relations between time points and/or graph elements may also be specified.<br><br>? $t_2, g_2, y_1, y_2$:<br>$\boldsymbol{f}(\mathbf{t_1}, \mathbf{g_1}) = y_1; \boldsymbol{f}(t_2, g_2) = y_2$;<br>$y_1 \boldsymbol{\Lambda} y_2$<br><br>Or<br><br>Find attribute values related in the given way where one of the values occurs at a given time, and the other is associated with a given graph element. Also find the unspecified graph element and time point associated with the attribute values. Relations between time points and/or graph elements may also be specified.<br><br>? $t_2, g_1, y_1, y_2$:<br>$\boldsymbol{f}(\mathbf{t_1}, g_1) = y_1; \boldsymbol{f}(t_2, \mathbf{g_2}) = y_2$;<br>$t_1 \boldsymbol{\Psi} t_2$; $g_1 \boldsymbol{\Phi} g_2$; $y_1 \boldsymbol{\Lambda} y_2$ | Find attribute values related in the given way where one of the values occurs at the given time. Relations between time points and graph elements may also be specified.<br><br>$t_2, g_1, g_2, y_1, y_2$:<br>$\boldsymbol{f}(\mathbf{t_1}, g_1) = y_1; \boldsymbol{f}(t_2, g_2) = y_2$;<br>$\mathbf{t_1} \boldsymbol{\Psi} t_2$; $g_1 \boldsymbol{\Phi} g_2$; $y_1 \boldsymbol{\Lambda} y_2$ |

| | | Graph elements (nodes, edges, graph objects) | | | |
|---|---|---|---|---|---|
| | | Both constraints | | One constraint, one target | Both are targets |
| | | Same element | Different elements | | |
| | Both are targets | Find attribute values (and the time points at which they occur) associated with the same given graph element, which are related in the given way. A relation between time points may also be specified.<br><br>? $t_1$, $t_2$, $y_1$, $y_2$:<br>$\boldsymbol{f}(t_1, \mathbf{g}) = y_1$; $\boldsymbol{f}(t_2, \mathbf{g}) = y_2$;<br>$y_1 \boldsymbol{\Lambda} y_2$; $t_1 \boldsymbol{\Psi} t_2$; | Find attribute values (and the time points at which they occur) associated with two given graph elements, which are related in the given way. A relation between time points may also be specified.<br><br>? $t_1$, $t_2$, $y_1$, $y_2$:<br>$\boldsymbol{f}(t_1, \mathbf{g_1}) = y_1$; $\boldsymbol{f}(t_2, \mathbf{g_2}) = y_2$;<br>$t_1 \boldsymbol{\Psi} t_2$; $y_1 \boldsymbol{\Lambda} y_2$ | Find attribute values related in the given way, where one of the attribute values is associated with a given graph element. Relations between time points and/or graph elements may also be specified.<br><br>? $t_1$, $t_2$, $g_2$, $y_1$, $y_2$:<br>$\boldsymbol{f}(t_1, \mathbf{g_1}) = y_1$; $\boldsymbol{f}(t_2, g_2) = y_2$;<br>$t_1 \boldsymbol{\Psi} t_2$; $g_1 \boldsymbol{\Phi} g_2$; $y_1 \boldsymbol{\Lambda} y_2$ | Find attribute values related in the given way. Relations between time points and/or graph elements may also be specified.<br><br>? $t_1$, $t_2$, $g_1$, $g_2$, $y_1$, $y_2$:<br>$\boldsymbol{f}(t_1, g_1) = y_1$; $\boldsymbol{f}(t_2, g_2) = y_2$;<br>$t_1 \boldsymbol{\Psi} t_2$; $g_1 \boldsymbol{\Phi} g_2$; $y_1 \boldsymbol{\Lambda} y_2$ |

**Figure 11 Relation seeking quadrant 2: considers synoptic relation seeking involving the behaviour of an attribute over the graph (or a graph subset)**

<table>
<tr><th colspan="3"></th><th colspan="4">Graph subsets</th></tr>
<tr><th colspan="3"></th><th colspan="2">Both constraints</th><th rowspan="2">One constraint, one target</th><th rowspan="2">Both are targets</th></tr>
<tr><th colspan="3"></th><th>Same subset</th><th>Different subsets</th></tr>
<tr><th rowspan="3">Time points</th><th rowspan="2">Both constraints</th><th>Same time</th><td colspan="2" rowspan="2">Not applicable</td><td>Find a pattern and the graph subset over which it occurs at a given time point, which is related in the given way to a pattern over a given graph subset at the same time point. A relation between graph subsets may also be specified.<br><br>? $G''$, $P_1$, $P_2$ :<br>$\boldsymbol{\beta}(\boldsymbol{f}(x_1, x_2) \mid x_1 \in \mathbf{G'}, x_2 = \mathbf{t}) \approx P_1$;<br>$\boldsymbol{\beta}(\boldsymbol{f}(x_1, x_2) \mid x_1 \in G'', x_2 = \mathbf{t}) \approx P_2$;<br>$\mathbf{G'}\ \boldsymbol{\Phi}\ G''$;<br>$P_1 \boldsymbol{\Lambda} P_2$</td><td>Find patterns related in the given way at the same time point. A relation between graph subsets may also be specified.<br><br>? $G'$, $G''$, $P_1$, $P_2$ :<br>$\boldsymbol{\beta}(\boldsymbol{f}(x_1, x_2) \mid x_1 \in G', x_2 = \mathbf{t}) \approx P_1$;<br>$\boldsymbol{\beta}(\boldsymbol{f}(x_1, x_2) \mid x_1 \in G'', x_2 = \mathbf{t}) \approx P_2$;<br>$G'\ \boldsymbol{\Phi}\ G''$;<br>$P_1 \boldsymbol{\Lambda} P_2$</td></tr>
<tr><th>Different times</th><td>Tasks as above, but involving two different time points.<br><br>? $G''$, $P_1$, $P_2$ :<br>$\boldsymbol{\beta}(\boldsymbol{f}(x_1, x_2) \mid x_1 \in \mathbf{G'}, x_2 = \mathbf{t_1}) \approx P_1$;<br>$\boldsymbol{\beta}(\boldsymbol{f}(x_1, x_2) \mid x_1 \in G'', x_2 = \mathbf{t_2}) \approx P_2$;<br>$\mathbf{G'}\ \boldsymbol{\Phi}\ G''$;<br>$P_1 \boldsymbol{\Lambda} P_2$</td><td>Tasks as above, but involving two different time points.<br><br>? $G'$, $G''$, $P_1$, $P_2$ :<br>$\boldsymbol{\beta}(\boldsymbol{f}(x_1, x_2) \mid x_1 \in G', x_2 = \mathbf{t_1}) \approx P_1$;<br>$\boldsymbol{\beta}(\boldsymbol{f}(x_1, x_2) \mid x_1 \in G'', x_2 = \mathbf{t_2}) \approx P_2$;<br>$\mathbf{G'}\ \boldsymbol{\Phi}\ G''$;<br>$P_1 \boldsymbol{\Lambda} P_2$</td></tr>
<tr><th colspan="2">One constraint, one target</th><td>Find a pattern (and time point) associated with a given graph subset, which is related in the given way to a pattern associated with the same graph subset at a given time. A relation between time points may also be specified.<br><br>? $t_2$, $P_1$, $P_2$ :</td><td>Find a pattern (and time point) associated with a given graph subset, which is related in the given way to a pattern associated with a different given graph subset at a given time. A relation between time points may also be specified.<br><br>? $t_2$, $P_1$, $P_2$ :</td><td>Find a pattern and the time point and graph subset over which it occurs related in the given way to a pattern associated with a given graph subset at a given time point. Relations between time points and graph subsets may also be specified.<br><br>? $G''$, $t_2$, $P_1$, $P_2$ :<br>$\boldsymbol{\beta}(\boldsymbol{f}(x_1, x_2) \mid x_1 \in \mathbf{G'}, x_2 = \mathbf{t_1}) \approx P_1$;<br>$\boldsymbol{\beta}(\boldsymbol{f}(x_1, x_2) \mid x_1 \in G'', x_2 = t_2) \approx P_2$;</td><td>Find patterns related in the given way where one of the patterns occurs at the given time. Relations between time points and graph subsets may also be specified.<br><br>? $G'$, $G''$, $t_1$, $P_1$, $P_2$ :<br>$\boldsymbol{\beta}(\boldsymbol{f}(x_1, x_2) \mid x_1 \in G', x_2 = \mathbf{t_1}) \approx P_1$;<br>$\boldsymbol{\beta}(\boldsymbol{f}(x_1, x_2) \mid x_1 \in G'', x_2 = t_2) \approx P_2$;<br>$\mathbf{t_1}\ \boldsymbol{\Psi}\ t_2$;<br>$G'\ \boldsymbol{\Phi}\ G''$;</td></tr>
</table>

| | | Graph subsets | | | |
|---|---|---|---|---|---|
| | | Both constraints | | One constraint, one target | Both are targets |
| | | Same subset | Different subsets | | |
| | | $\boldsymbol{\beta}(\boldsymbol{f}(x_1, x_2) \mid x_1 \in \mathbf{G'}, x_2 = \mathbf{t_1}) \approx P_1$;<br>$\boldsymbol{\beta}(\boldsymbol{f}(x_1, x_2) \mid x_1 \in \mathbf{G'}, x_2 = t_2) \approx P_2$;<br>$\mathbf{t_1}\ \boldsymbol{\Psi}\ t_2$;<br>$P_1\ \boldsymbol{\Lambda}\ P_2$ | $\boldsymbol{\beta}(\boldsymbol{f}(x_1, x_2) \mid x_1 \in \mathbf{G'}, x_2 = \mathbf{t_1}) \approx P_1$;<br>$\boldsymbol{\beta}(\boldsymbol{f}(x_1, x_2) \mid x_1 \in \mathbf{G''}, x_2 = t_2) \approx P_2$;<br>$\mathbf{t_1}\ \boldsymbol{\Psi}\ t_2$;<br>$P_1\ \boldsymbol{\Lambda}\ P_2$ | $\mathbf{t_1}\ \boldsymbol{\Psi}\ t_2$;<br>$\mathbf{G'}\ \boldsymbol{\Phi}\ G''$;<br>$P_1\ \boldsymbol{\Lambda}\ P_2$<br><br>Or<br><br>Find patterns related in the given way where one of the patterns occurs at a given time, and the other occurs over a given graph subset. Also find the unspecified graph subset and time point over which/at the patterns occur. Relations between time points and graph subsets may also be specified.<br><br>? = $G''$, $t_1$, $P_1$, $P_2$ :<br>$\boldsymbol{\beta}(\boldsymbol{f}(x_1, x_2) \mid x_1 \in \mathbf{G'}, x_2 = t_1) \approx P_1$;<br>$\boldsymbol{\beta}(\boldsymbol{f}(x_1, x_2) \mid x_1 \in G'', x_2 = \mathbf{t_2}) \approx P_2$;<br>$t_1\ \boldsymbol{\Psi}\ \mathbf{t_2}$;<br>$\mathbf{G'}\ \boldsymbol{\Phi}\ G''$;<br>$P_1\ \boldsymbol{\Lambda}\ P_2$ | $P_1\ \boldsymbol{\Lambda}\ P_2$ |
| | Both are targets | Find patterns (and the time points at which they occur) associated with a single given graph subset, which are related in the given way. A relation between time points may also be specified.<br>? $t_1$, $t_2$, $P_1$, $P_2$ :<br>$\boldsymbol{\beta}(\boldsymbol{f}(x_1, x_2) \mid x_1 \in \mathbf{G'}, x_2 = t_1) \approx P_1$;<br>$\boldsymbol{\beta}(\boldsymbol{f}(x_1, x_2) \mid x_1 \in \mathbf{G'}, x_2 = t_2) \approx P_2$;<br>$t_1\ \boldsymbol{\Psi}\ t_2$;<br>$P_1\ \boldsymbol{\Lambda}\ P_2$ | Find patterns (and the time points at which they occur) associated with two given graph subsets, which are related in the given way. A relation between time points may also be specified.<br>? $t_1$, $t_2$, $P_1$, $P_2$ :<br>$\boldsymbol{\beta}(\boldsymbol{f}(x_1, x_2) \mid x_1 \in \mathbf{G'}, x_2 = t_1) \approx P_1$;<br>$\boldsymbol{\beta}(\boldsymbol{f}(x_1, x_2) \mid x_1 \in \mathbf{G''}, x_2 = t_2) \approx P_2$;<br>$t_1\ \boldsymbol{\Psi}\ t_2$;<br>$P_1\ \boldsymbol{\Lambda}\ P_2$ | Find patterns related in the given way, where one of the patterns is associated with a given graph subset. Relations between time points and graph subsets may also be specified.<br><br>? $G''$, $t_1$, $t_2$, $P_1$, $P_2$ :<br>$\boldsymbol{\beta}(\boldsymbol{f}(x_1, x_2) \mid x_1 \in \mathbf{G'}, x_2 = t_1) \approx P_1$;<br>$\boldsymbol{\beta}(\boldsymbol{f}(x_1, x_2) \mid x_1 \in G'', x_2 = t_2) \approx P_2$;<br>$t_1\ \boldsymbol{\Psi}\ t_2$;<br>$\mathbf{G'}\ \boldsymbol{\Phi}\ G''$;<br>$P_1\ \boldsymbol{\Lambda}\ P_2$ | Find patterns related in the given way. Relations between time points and graph subsets may also be specified.<br><br>? $G'$, $G''$, $t_1$, $t_2$, $P_1$, $P_2$ :<br>$\boldsymbol{\beta}(\boldsymbol{f}(x_1, x_2) \mid x_1 \in G', x_2 = t_1) \approx P_1$;<br>$\boldsymbol{\beta}(\boldsymbol{f}(x_1, x_2) \mid x_1 \in G'', x_2 = t_2) \approx P_2$;<br>$t_1\ \boldsymbol{\Psi}\ t_2$;<br>$G'\ \boldsymbol{\Phi}\ G''$;<br>$P_1\ \boldsymbol{\Lambda}\ P_2$ |

**Figure 12 Relation seeking quadrant 3: considers relation seeking tasks involving the behaviour of an attribute of a single graph element over time (i.e. a temporal trend)**

<table>
<tr><th colspan="3"></th><th colspan="4">Graph elements (nodes, edges, graph objects)</th></tr>
<tr><th colspan="3"></th><th colspan="2">Both constraints</th><th rowspan="2">One constraint, one target</th><th rowspan="2">Both are targets</th></tr>
<tr><th colspan="3"></th><th>Same element</th><th>Different elements</th></tr>
<tr><td rowspan="2">Time intervals</td><td rowspan="2">Both constraints</td><td>Same time</td><td colspan="2" rowspan="2">Not applicable</td><td>Find a pattern (and the graph element associated with it) which occurs over a given time interval and is related in the given way to a pattern associated with a given graph element over the same time interval. A relation between graph elements may also be specified.[16]<br><br>? $g_2$, $P_1$, $P_2$:<br>$\boldsymbol{\beta}(\boldsymbol{f}(x_1, x_2) \mid x_1 = \mathbf{g_1}, x_2 \in \mathbf{T'}) \approx P_1$;<br>$\boldsymbol{\beta}(\boldsymbol{f}(x_1, x_2) \mid x_1 = g_2, x_2 \in \mathbf{T'}) \approx P_2$;<br>$P_1 \wedge P_2$;<br>$\mathbf{g_1}\, \Phi\, g_2$</td><td>Find patterns (and their associated graph elements) which occur over the same given time interval and are related in the given way. A relation between graph elements may also be specified.<br><br>? $g_1$, $g_2$, $P_1$, $P_2$:<br>$\boldsymbol{\beta}(\boldsymbol{f}(x_1, x_2) \mid x_1 = g_1, x_2 \in \mathbf{T'}) \approx P_1$;<br>$\boldsymbol{\beta}(\boldsymbol{f}(x_1, x_2) \mid x_1 = g_2, x_2 \in \mathbf{T'}) \approx P_2$;<br>$P_1 \wedge P_2$;<br>$g_1\, \Phi\, g_2$</td></tr>
<tr><td>Different times</td><td>Find a pattern (and the graph element associated with it) which occurs over a given time interval and is related in the given way to a pattern associated with a given graph element over a given time interval. A relation between graph elements may also be specified<br><br>? $g_2$, $P_1$, $P_2$:<br>$\boldsymbol{\beta}(\boldsymbol{f}(x_1, x_2) \mid x_1 = \mathbf{g_1}, x_2 \in \mathbf{T'}) \approx P_1$;<br>$\boldsymbol{\beta}(\boldsymbol{f}(x_1, x_2) \mid x_1 = g_2, x_2 \in \mathbf{T'}) \approx P_2$;<br>$P_1 \wedge P_2$;<br>$\mathbf{g_1}\, \Phi\, g_2$</td><td>Find patterns (and their associated graph elements) which occur over two given time intervals and are related in the given way. A relation between graph elements may also be specified<br><br>? $g_1$, $g_2$, $P_1$, $P_2$:<br>$\boldsymbol{\beta}(\boldsymbol{f}(x_1, x_2) \mid x_1 = g_1, x_2 \in \mathbf{T'}) \approx P_1$;<br>$\boldsymbol{\beta}(\boldsymbol{f}(x_1, x_2) \mid x_1 = g_2, x_2 \in \mathbf{T''}) \approx P_2$;<br>$P_1 \wedge P_2$;<br>$g_1\, \Phi\, g_2$</td></tr>
</table>

[16] In all cases in this table, if we wish to specify an association relation between the graph elements, we must also specify a time at which the association relation occurs i.e. ($\mathbf{g_1}$, t) $\Phi$ ($g_2$, t): *'a given association relation exists between the graph elements at time* **t***'.*

| | Graph elements (nodes, edges, graph objects) | | | |
|---|---|---|---|---|
| | Both constraints | | One constraint, one target | Both are targets |
| | Same element | Different elements | | |
| One constraint, one target | Find a pattern (and the time interval over which it occurs) for a given graph element, which is related in the given way to a pattern associated with the same graph element over a given time interval. A relation between time intervals may also be specified.<br><br>? $T''$, $P_1$, $P_2$:<br>$\boldsymbol{\beta}(\boldsymbol{f}(x_1, x_2) \mid x_1 = \mathbf{g}, x_2 \in \mathbf{T'}) \approx P_1$;<br>$\boldsymbol{\beta}(\boldsymbol{f}(x_1, x_2) \mid x_1 = \mathbf{g}, x_2 \in T'') \approx P_2$;<br>$\mathbf{T'}\ \boldsymbol{\Psi}\ T''$;<br>$P_1 \ \boldsymbol{\Lambda}\ P_2$ | Find a pattern (and the time interval over which it occurs) for a given graph element, which is related in the given way to a pattern associated with a given graph element over a given time interval. A relation between time intervals may also be specified.<br><br>? $T''$, $P_1$, $P_2$:<br>$\boldsymbol{\beta}(\boldsymbol{f}(x_1, x_2) \mid x_1 = \mathbf{g_1}, x_2 \in \mathbf{T'}) \approx P_1$;<br>$\boldsymbol{\beta}(\boldsymbol{f}(x_1, x_2) \mid x_1 = \mathbf{g_2}, x_2 \in T'') \approx P_2$;<br>$T'\ \boldsymbol{\Psi}\ T''$;<br>$P_1 \ \boldsymbol{\Lambda}\ P_2$ | Find a pattern, and the graph element and time interval over which it occurs, which is related in the given way to a pattern associated with a given graph element over a given time interval. A relation between time intervals and/or graph elements may also be specified.<br><br>? $g_2$, $T''$, $P_1$, $P_2$:<br>$\boldsymbol{\beta}(\boldsymbol{f}(x_1, x_2) \mid x_1 = \mathbf{g_1}, x_2 \in \mathbf{T'}) \approx P_1$;<br>$\boldsymbol{\beta}(\boldsymbol{f}(x_1, x_2) \mid x_1 = g_2, x_2 \in T'') \approx P_2$;<br>$\mathbf{T'}\ \boldsymbol{\Psi}\ T''$;<br>$P_1 \ \boldsymbol{\Lambda}\ P_2$;<br>$\mathbf{g_1}\ \Phi\ g_2$<br><br>Or<br><br>Find patterns related in the given way where one of the patterns occurs over a given time interval, and the other is associated with a given graph element. Also find the unspecified graph element and time interval associated with the patterns). A relation between time intervals and/or graph elements may also be specified.<br><br>? $g_2$, $T'$, $P_1$, $P_2$:<br>$\boldsymbol{\beta}(\boldsymbol{f}(x_1, x_2) \mid x_1 = \mathbf{g_1}, x_2 \in T') \approx P_1$;<br>$\boldsymbol{\beta}(\boldsymbol{f}(x_1, x_2) \mid x_1 = g_2, x_2 \in \mathbf{T''}) \approx P_2$;<br>$T'\ \boldsymbol{\Psi}\ \mathbf{T''}$;<br>$P_1 \ \boldsymbol{\Lambda}\ P_2$;<br>$\mathbf{g_1}\ \Phi\ g_2$ | Find patterns related in the given way where one of the patterns occurs over a given time interval. A relation between time intervals and/or graph elements may also be specified.<br><br>? $g_1$, $g_2$, $T''$, $P_1$, $P_2$:<br>$\boldsymbol{\beta}(\boldsymbol{f}(x_1, x_2) \mid x_1 = g_1, x_2 \in \mathbf{T'}) \approx P_1$;<br>$\boldsymbol{\beta}(\boldsymbol{f}(x_1, x_2) \mid x_1 = g_2, x_2 \in T'') \approx P_2$;<br>$\mathbf{T'}\ \boldsymbol{\Psi}\ T''$;<br>$P_1 \ \boldsymbol{\Lambda}\ P_2$;<br>$g_1\ \Phi\ g_2$ |

| | | Graph elements (nodes, edges, graph objects) | | | |
|---|---|---|---|---|---|
| | | Both constraints | | One constraint, one target | Both are targets |
| | | Same element | Different elements | | |
| | Both are targets | Find patterns (and the time intervals over which they occur) associated with a single graph element, which are related in the given way. A relation between time intervals may also be specified.<br><br>? $T', T'', P_1, P_2$:<br>$\boldsymbol{\beta}(\boldsymbol{f}(x_1, x_2) \mid x_1 = \mathbf{g}, x_2 \in T') \approx P_1$;<br>$\boldsymbol{\beta}(\boldsymbol{f}(x_1, x_2) \mid x_1 = \mathbf{g}, x_2 \in T'') \approx P_2$;<br>$T' \,\boldsymbol{\Psi}\, T''$;<br>$P_1 \wedge P_2$ | Find patterns (and the time intervals over which they occur) associated with two given graph elements, which are related in the given way. A relation between time intervals may also be specified.<br><br>? $T', T'', P_1, P_2$:<br>$\boldsymbol{\beta}(\boldsymbol{f}(x_1, x_2) \mid x_1 = \mathbf{g_1}, x_2 \in T') \approx P_1$;<br>$\boldsymbol{\beta}(\boldsymbol{f}(x_1, x_2) \mid x_1 = \mathbf{g_2}, x_2 \in T'') \approx P_2$;<br>$T' \,\boldsymbol{\Psi}\, T''$;<br>$P_1 \wedge P_2$ | Find patterns related in the given way, where one of the patterns is associated with a given graph element. A relation between time intervals and/or graph elements may also be specified.<br><br>$g_2, T', T'', P_1, P_2$:<br>$\boldsymbol{\beta}(\boldsymbol{f}(x_1, x_2) \mid x_1 = \mathbf{g_1}, x_2 \in T') \approx P_1$;<br>$\boldsymbol{\beta}(\boldsymbol{f}(x_1, x_2) \mid x_1 = g_2, x_2 \in T'') \approx P_2$;<br>$T' \,\boldsymbol{\Psi}\, T''$;<br>$P_1 \wedge P_2$;<br>$\mathbf{g_1} \,\Phi\, g_2$ | Find patterns related in the given way. A relation between time intervals and/or graph elements may also be specified.<br><br>? $g_1, g_2, T', T'', P_1, P_2$:<br>$\boldsymbol{\beta}(\boldsymbol{f}(x_1, x_2) \mid x_1 = g_1, x_2 \in T') \approx P_1$;<br>$\boldsymbol{\beta}(\boldsymbol{f}(x_1, x_2) \mid x_1 = g_2, x_2 \in T'') \approx P_2$;<br>$T' \,\boldsymbol{\Psi}\, T''$;<br>$P_1 \wedge P_2$;<br>$g_1 \,\Phi\, g_2$ |

**Figure 13 Relation seeking quadrant 4: considers relation seeking tasks involving aspectual behaviours (i) the behaviour of temporal trends for all graph elements, over the graph (ii) the behaviour of an attribute over the graph, over time**

| | Graph subsets: Both constraints – Same subset | Graph subsets: Both constraints – Different subsets | Graph subsets: One constraint, one target | Graph subsets: Both are targets |
|---|---|---|---|---|
| **Time intervals – Both constraints – Same time** | Not applicable | Not applicable | Find a pattern (and the graph subset associated with it) which is associated with a given time interval and is related in the given way to a pattern associated with a given graph subset and the same time interval. A relation between graph subsets may also be specified.<br><br>$?\ G'', P_1, P_2$:<br>$\boldsymbol{\beta}_G\{\boldsymbol{\beta}_T[\boldsymbol{f}(x_1, x_2) \mid x_2 \in \mathbf{T'})] \mid x_1 \in \mathbf{G'}\} \approx P_1$;<br>$\boldsymbol{\beta}_G\{\boldsymbol{\beta}_T[\boldsymbol{f}(x_1, x_2) \mid x_2 \in \mathbf{T'})] \mid x_1 \in G''\} \approx P_2$;<br>$\mathbf{G'}\ \boldsymbol{\Phi}\ G''$;<br>$P_1 \wedge P_2$<br><br>or<br><br>$?\ G'', P_1, P_2$:<br>$\boldsymbol{\beta}_T\{\boldsymbol{\beta}_G[\boldsymbol{f}(x_1, x_2) \mid x_1 \in \mathbf{G'})] \mid x_2 \in \mathbf{T'}\} \approx P_1$; $\boldsymbol{\beta}_T\{\boldsymbol{\beta}_G[\boldsymbol{f}(x_1, x_2) \mid x_1 \in G'')] \mid x_2 \in \mathbf{T'}\} \approx P_2$;<br>$\mathbf{G'}\ \boldsymbol{\Phi}\ G''$;<br>$P_1 \wedge P_2$ | Find patterns (and their associated graph subsets) which are associated with a single given time interval and are related in the given way. A relation between graph subsets may also be specified.<br><br>$?\ G', G'', P_1, P_2$:<br>$\boldsymbol{\beta}_G\{\boldsymbol{\beta}_T[\boldsymbol{f}(x_1, x_2) \mid x_2 \in \mathbf{T'})] \mid x_1 \in G'\} \approx P_1$;<br>$\boldsymbol{\beta}_G\{\boldsymbol{\beta}_T[\boldsymbol{f}(x_1, x_2) \mid x_2 \in \mathbf{T'})] \mid x_1 \in G''\} \approx P_2$;<br>$G'\ \boldsymbol{\Phi}\ G''$;<br>$P_1 \wedge P_2$<br><br>or<br><br>$?\ G', G'', P_1, P_2$:<br>$\boldsymbol{\beta}_T\{\boldsymbol{\beta}_G[\boldsymbol{f}(x_1, x_2) \mid x_1 \in G')] \mid x_2 \in \mathbf{T'}\} \approx P_1$;<br>$\boldsymbol{\beta}_T\{\boldsymbol{\beta}_G[\boldsymbol{f}(x_1, x_2) \mid x_1 \in G'')] \mid x_2 \in \mathbf{T'}\} \approx P_2$;<br>$G'\ \boldsymbol{\Phi}\ G''$;<br>$P_1 \wedge P_2$ |
| **Time intervals – Both constraints – Different times** | Not applicable | Not applicable | Find a pattern (and the graph subset associated with it) which is associated with a given time interval and is related in the given way to a pattern associated with a given graph subset and a given time interval. A relation between graph subsets may also be specified.<br><br>$?\ G'', P_1, P_2$:<br>$\boldsymbol{\beta}_G\{\boldsymbol{\beta}_T[\boldsymbol{f}(x_1, x_2) \mid x_2 \in \mathbf{T'})] \mid x_1 \in \mathbf{G'}\} \approx P_1$;<br>$\boldsymbol{\beta}_G\{\boldsymbol{\beta}_T[\boldsymbol{f}(x_1, x_2) \mid x_2 \in \mathbf{T''})] \mid x_1 \in G''\} \approx P_2$; | Find patterns (and their associated graph subsets) which are associated with two given time intervals and are related in the given way. A relation between graph subsets may also be specified.<br><br>$?\ G', G'', P_1, P_2$:<br>$\boldsymbol{\beta}_G\{\boldsymbol{\beta}_T[\boldsymbol{f}(x_1, x_2) \mid x_2 \in \mathbf{T'})] \mid x_1 \in G'\} \approx P_1$;<br>$\boldsymbol{\beta}_G\{\boldsymbol{\beta}_T[\boldsymbol{f}(x_1, x_2) \mid x_2 \in \mathbf{T''})] \mid x_1 \in G''\} \approx P_2$;<br>$G'\ \boldsymbol{\Phi}\ G''$; |

| | Graph subsets | | | |
|---|---|---|---|---|
| | Both constraints | | One constraint, one target | Both are targets |
| | Same subset | Different subsets | | |
| | | | $\mathbf{G'}\ \Phi\ G''$;<br>$P_1 \wedge P_2$<br><br>or<br><br>? $G''$, $P_1$, $P_2$:<br>$\beta_T\{\beta_G[f(x_1, x_2) \mid x_1 \in \mathbf{G'})] \mid x_2 \in \mathbf{T'}\} \approx P_1$; $\beta_T\{\beta_G[f(x_1, x_2) \mid x_1 \in G'')] \mid x_2 \in \mathbf{T''}\} \approx P_2$;<br>$\mathbf{G'}\ \Phi\ G''$;<br>$P_1 \wedge P_2$ | $P_1 \wedge P_2$<br><br>or<br><br>? $G'$, $G''$, $P_1$, $P_2$:<br>$\beta_T\{\beta_G[f(x_1, x_2) \mid x_1 \in G')] \mid x_2 \in \mathbf{T'}\} \approx P_1$;<br>$\beta_T\{\beta_G[f(x_1, x_2) \mid x_1 \in G'')] \mid x_2 \in \mathbf{T''}\} \approx P_2$;<br>$G'\ \Phi\ G''$;<br>$P_1 \wedge P_2$ |
| One constraint, one target | Find a pattern (and the time interval with which it is associated) for a given graph subset, which is related in the given way to a pattern associated with the same graph subset and a given time interval. A relation between time intervals may also be specified.<br><br>? $T''$, $P_1$, $P_2$:<br>$\beta_G\{\beta_T[f(x_1, x_2) \mid x_2 \in \mathbf{T'})] \mid x_1 \in \mathbf{G'}\} \approx P_1$;<br>$\beta_G\{\beta_T[f(x_1, x_2) \mid x_2 \in T'')] \mid x_1 \in \mathbf{G'}\} \approx P_2$;<br>$\mathbf{T'}\ \Psi\ T''$;<br>$P_1 \wedge P_2$<br><br>or<br><br>? $T''$, $P_1$, $P_2$:<br>$\beta_T\{\beta_G[f(x_1, x_2) \mid x_1 \in \mathbf{G'})] \mid x_2 \in \mathbf{T'}\} \approx P_1$;<br>$\beta_T\{\beta_G[f(x_1, x_2) \mid x_1 \in \mathbf{G'})] \mid x_2 \in T''\} \approx P_2$;<br>$\mathbf{T'}\ \Psi\ T''$;<br>$P_1 \wedge P_2$ | Find patterns (and the time intervals over which they occur) associated with two given graph subsets, which are related in the given way. A relation between time intervals may also be specified.<br><br>? $T''$, $P_1$, $P_2$:<br>$\beta_G\{\beta_T[f(x_1, x_2) \mid x_2 \in \mathbf{T'})] \mid x_1 \in \mathbf{G'}\} \approx P_1$;<br>$\beta_G\{\beta_T[f(x_1, x_2) \mid x_2 \in T'')] \mid x_1 \in \mathbf{G''}\} \approx P_2$;<br>$\mathbf{T'}\ \Psi\ T''$;<br>$P_1 \wedge P_2$<br><br>or<br><br>? $T''$, $P_1$, $P_2$:<br>$\beta_T\{\beta_G[f(x_1, x_2) \mid x_1 \in \mathbf{G'})] \mid x_2 \in \mathbf{T'}\} \approx P_1$;<br>$\beta_T\{\beta_G[f(x_1, x_2) \mid x_1 \in \mathbf{G''})] \mid x_2 \in T''\} \approx P_2$;<br>$T'\ \Psi\ T''$;<br>$P_1 \wedge P_2$ | Find a pattern, and the graph subset and time interval with which it is associated, which is related in the given way to a pattern associated with a given graph subset and a given time interval. Relations between time intervals and/or graph subsets may also be specified.<br><br>? $G''$, $T''$, $P_1$, $P_2$:<br>$\beta_G\{\beta_T[f(x_1, x_2) \mid x_2 \in \mathbf{T'})] \mid x_1 \in \mathbf{G'}\} \approx P_1$;<br>$\beta_G\{\beta_T[f(x_1, x_2) \mid x_2 \in T'')] \mid x_1 \in G''\} \approx P_2$;<br>$\mathbf{T'}\ \Psi\ T''$;<br>$\mathbf{G'}\ \Phi\ G''$;<br>$P_1 \wedge P_2$<br><br>or<br><br>? $G''$, $T''$, $P_1$, $P_2$:<br>$\beta_T\{\beta_G[f(x_1, x_2) \mid x_1 \in \mathbf{G'})] \mid x_2 \in \mathbf{T'}\} \approx P_1$; $\beta_T\{\beta_G[f(x_1, x_2) \mid x_1 \in G'')] \mid x_2 \in T''\} \approx P_2$;<br>$\mathbf{T'}\ \Psi\ T''$;<br>$\mathbf{G'}\ \Phi\ G''$;<br>$P_1 \wedge P_2$ | Find patterns related in the given way where one of the patterns involves a given time interval. Relations between time intervals and/or graph subsets may also be specified.<br><br>? $G'$, $G''$, $T''$, $P_1$, $P_2$:<br>$\beta_G\{\beta_T[f(x_1, x_2) \mid x_2 \in \mathbf{T'})] \mid x_1 \in G'\} \approx P_1$;<br>$\beta_G\{\beta_T[f(x_1, x_2) \mid x_2 \in T'')] \mid x_1 \in G''\} \approx P_2$;<br>$\mathbf{T'}\ \Psi\ T''$;<br>$G'\ \Phi\ G''$;<br>$P_1 \wedge P_2$<br><br>or<br><br>? $G'$, $G''$, $T''$, $P_1$, $P_2$:<br>$\beta_T\{\beta_G[f(x_1, x_2) \mid x_1 \in G')] \mid x_2 \in \mathbf{T'}\} \approx P_1$;<br>$\beta_T\{\beta_G[f(x_1, x_2) \mid x_1 \in G'')] \mid x_2 \in T''\} \approx P_2$;<br>$\mathbf{T'}\ \Psi\ T''$;<br>$G'\ \Phi\ G''$;<br>$P_1 \wedge P_2$ |

| | | Graph subsets | | | |
|---|---|---|---|---|---|
| | | Both constraints | | One constraint, one target | Both are targets |
| | | Same subset | Different subsets | | |
| | | | | OR<br><br>Find patterns related in the given way where one of the patterns is associated with a given time interval, and the other is associated with a given graph subset. Also find the unspecified graph subset and time interval associated with the patterns). Relations between time intervals and/or graph subsets may also be specified.<br><br>$?\, G', T'', P_1, P_2$:<br>$\boldsymbol{\beta_G}\{\boldsymbol{\beta_T}[\boldsymbol{f}(x_1, x_2) \mid x_2 \in \mathbf{T'})] \mid x_1 \in G'\} \approx P_1$;<br>$\boldsymbol{\beta_G}\{\boldsymbol{\beta_T}[\boldsymbol{f}(x_1, x_2) \mid x_2 \in T'')] \mid x_1 \in \mathbf{G''}\} \approx P_2$;<br>$T'\ \boldsymbol{\Psi}\ \mathbf{T''}$;<br>$\mathbf{G'}\ \boldsymbol{\Phi}\ G''$;<br>$P_1 \boldsymbol{\Lambda} P_2$<br><br>or<br><br>$?\, G'', T', T'', P_1, P_2$:<br>$\boldsymbol{\beta_T}\{\boldsymbol{\beta_G}[\boldsymbol{f}(x_1, x_2) \mid x_1 \in \mathbf{G'})] \mid x_2 \in T'\} \approx P_1$; $\boldsymbol{\beta_T}\{\boldsymbol{\beta_G}[\boldsymbol{f}(x_1, x_2) \mid x_1 \in G'')] \mid x_2 \in \mathbf{T''}\} \approx P_2$;<br>$T'\ \boldsymbol{\Psi}\ \mathbf{T''}$;<br>$\mathbf{G'}\ \boldsymbol{\Phi}\ G''$;<br>$P_1 \boldsymbol{\Lambda} P_2$ | |
| | Both are targets | Find patterns (and the time intervals over which they occur) associated with the same given graph subset, which are related in the given way. A relation between time intervals may also be specified. | Find patterns (and the time intervals over which they occur) associated with two given graph subsets, which are related in the given way. A relation between time intervals may also be specified. | Find patterns related in the given way, where one pattern is associated with a given graph subset. Relations between time intervals and/or graph subsets may also be specified.<br><br>$?\, G'', T', T'', P_1, P_2$: | Find patterns related in the given way. Relations between time intervals and/or graph subsets may also be specified.<br><br>$?\, G', G'', T', T'', P_1, P_2$:<br>$\boldsymbol{\beta_G}\{\boldsymbol{\beta_T}[\boldsymbol{f}(x_1, x_2) \mid x_2 \in T')] \mid x_1 \in G'\} \approx P_1$; |

| | | Graph subsets | | | |
|---|---|---|---|---|---|
| | | **Both constraints** | | **One constraint, one target** | **Both are targets** |
| | | **Same subset** | **Different subsets** | | |
| | | ? $T', T'', P_1, P_2$:<br>$\boldsymbol{\beta}_G\{\boldsymbol{\beta}_T[\boldsymbol{f}(x_1, x_2) \mid x_2 \in T')]\Vert x_1 \in \mathbf{G'}\} \approx P_1$;<br>$\boldsymbol{\beta}_G\{\boldsymbol{\beta}_T[\boldsymbol{f}(x_1, x_2) \mid x_2 \in T'')]\Vert x_1 \in \mathbf{G'}\} \approx P_2$;<br>$T' \boldsymbol{\Psi} T''$;<br>$P_1 \boldsymbol{\Lambda} P_2$<br><br>or<br><br>? $T', T'', P_1, P_2$:<br>$\boldsymbol{\beta}_T\{\boldsymbol{\beta}_G[\boldsymbol{f}(x_1, x_2) \mid x_1 \in \mathbf{G'})]\Vert x_2 \in T'\} \approx P_1$;<br>$\boldsymbol{\beta}_T\{\boldsymbol{\beta}_G[\boldsymbol{f}(x_1, x_2) \mid x_1 \in \mathbf{G'})]\Vert x_2 \in T''\} \approx P_2$;<br>$T' \boldsymbol{\Psi} T''$;<br>$P_1 \boldsymbol{\Lambda} P_2$ | ? $T', T'', P_1, P_2$:<br>$\boldsymbol{\beta}_G\{\boldsymbol{\beta}_T[\boldsymbol{f}(x_1, x_2) \mid x_2 \in T')]\Vert x_1 \in \mathbf{G'}\} \approx P_1$;<br>$\boldsymbol{\beta}_G\{\boldsymbol{\beta}_T[\boldsymbol{f}(x_1, x_2) \mid x_2 \in T'')]\Vert x_1 \in \mathbf{G''}\} \approx P_2$;<br>$T' \boldsymbol{\Psi} T''$;<br>$P_1 \boldsymbol{\Lambda} P_2$<br><br>or<br><br>? $T', T'', P_1, P_2$:<br>$\boldsymbol{\beta}_T\{\boldsymbol{\beta}_G[\boldsymbol{f}(x_1, x_2) \mid x_1 \in \mathbf{G'})]\Vert x_2 \in T'\} \approx P_1$;<br>$\boldsymbol{\beta}_T\{\boldsymbol{\beta}_G[\boldsymbol{f}(x_1, x_2) \mid x_1 \in \mathbf{G''})]\Vert x_2 \in T''\} \approx P_2$;<br>$T' \boldsymbol{\Psi} T''$;<br>$P_1 \boldsymbol{\Lambda} P_2$ | $\boldsymbol{\beta}_G\{\boldsymbol{\beta}_T[\boldsymbol{f}(x_1, x_2) \mid x_2 \in T')]\Vert x_1 \in \mathbf{G'}\} \approx P_1$;<br>$\boldsymbol{\beta}_G\{\boldsymbol{\beta}_T[\boldsymbol{f}(x_1, x_2) \mid x_2 \in T'')]\Vert x_1 \in G''\} \approx P_2$;<br>$T' \boldsymbol{\Psi} T''$;<br>$\mathbf{G'} \boldsymbol{\Phi} G''$;<br>$P_1 \boldsymbol{\Lambda} P_2$<br><br>or<br><br>? $G'', T', T'', P_1, P_2$:<br>$\boldsymbol{\beta}_T\{\boldsymbol{\beta}_G[\boldsymbol{f}(x_1, x_2) \mid x_1 \in \boldsymbol{G'})]\Vert x_2 \in T'\} \approx P_1$; $\boldsymbol{\beta}_T\{\boldsymbol{\beta}_G[\boldsymbol{f}(x_1, x_2) \mid x_1 \in G'')]\Vert x_2 \in T''\} \approx P_2$;<br>$T' \boldsymbol{\Psi} T''$;<br>$\mathbf{G'} \boldsymbol{\Phi} G''$;<br>$P_1 \boldsymbol{\Lambda} P_2$ | $\boldsymbol{\beta}_G\{\boldsymbol{\beta}_T[\boldsymbol{f}(x_1, x_2) \mid x_2 \in T'')]\Vert x_1 \in G''\} \approx P_2$;<br>$T' \boldsymbol{\Psi} T''$;<br>$G' \boldsymbol{\Phi} G''$;<br>$P_1 \boldsymbol{\Lambda} P_2$<br><br>or<br><br>? $G', G'', T', T'', P_1, P_2$:<br>$\boldsymbol{\beta}_T\{\boldsymbol{\beta}_G[\boldsymbol{f}(x_1, x_2) \mid x_1 \in G')]\Vert x_2 \in T'\} \approx P_1$;<br>$\boldsymbol{\beta}_T\{\boldsymbol{\beta}_G[\boldsymbol{f}(x_1, x_2) \mid x_1 \in G'')]\Vert x_2 \in T''\} \approx P_2$;<br>$T' \boldsymbol{\Psi} T''$;<br>$G' \boldsymbol{\Phi} G''$;<br>$P_1 \boldsymbol{\Lambda} P_2$ |

## 4.2 Structural tasks

As for the attribute based tasks, we divide the structural task space based on the referential components involved (Figure 14). Quadrant 1 contains the elementary tasks, while the other three quadrants contain the synoptic tasks involving the partial and aspectual structural patterns.

**Figure 14 Structural tasks based on referential components involved**

| | Pair of nodes | Graph subsets |
|---|---|---|
| **Time Points** | **Q1 Elementary**<br>Tasks involving connection between a pair of nodes at a single time point (inverse comparison and relation seeking only). | **Q2 Synoptic**<br>Tasks involving the behaviour, or configuration, of association relations within a set of nodes at a single time e.g. clusters, cliques, motifs etc.: |
| **Time Intervals** | **Q3 Synoptic**<br>Tasks involving the behaviour of association relations between two graph objects over time e.g. the pattern of change in connectivity between two nodes over time. | **Q4 Synoptic**<br>Tasks involving:<br><br>The behaviour of the collection of patterns in (Q3) i.e. the aggregate pattern of all association relations between pairs of graph objects over time, or the distribution of individual temporal behaviours over the graph.<br><br>OR<br><br>The behaviour of the configurations of association relations over the set of nodes (i.e. (Q2)), over time. |

### 4.2.1 Elementary structural tasks

We distinguish two main elementary tasks: finding association relations (connections) between elements and finding elements connected in the given way. Note that association relations may exist between individual graph elements (nodes) or subsets of the graph treated as individual elements (graph objects). We may also wish to compare and find relations between association relations. We discuss these tasks below.

#### *4.2.1.1 Find connections between elements (comparison)*

This is the "comparison" subtask: the term is rather unintuitive when dealing with questions of connectivity between graph objects, however, its definition of '*finding the relation between given elements*' is exactly what we are attempting to do here. This reflects scheme 1 of Andrienko's pure relational tasks ([2] p.62-63) (how are the elements **p** and **q** (or the subsets **P** and **Q**) of the set **S** related?):

(How) is $\mathbf{g_1}$ connected to $\mathbf{g_2}$ at the given time, **t**?

$$?\lambda: (\mathbf{g_1}, \mathbf{t})\ \lambda\ (\mathbf{g_2}, \mathbf{t})$$

e.g. *(how) are graph objects a and b connected at time 4?*

#### *4.2.1.2 Find elements connected in the given way (relation seeking)*

This is the relation seeking subtask, and there are two versions reflecting Andrienko's pure relational schemes 2 and 3, plus a third, hybrid version. Temporal variations of each task are also given in the summary table, below. The association relation can be specified as generally as stating whether or not any connection exists between two objects, or as specifically as including the distance, direction, and/or domain attributes.

(i) where the connection and one of the graph objects is specified (i.e. scheme 2: what element (or subset) of the set **S** is related to the element **p** (or subset **P**) in the way **ρ**?)

Find the graph object(s) to which graph object $\mathbf{g_1}$ is connected in the given way at time **t**:

$? \, g_2 : (\mathbf{g_1}, \mathbf{t}) \wedge (g_2, \mathbf{t})$

*e.g. find the node(s) directly connected to node a at time $t_4$; find node(s)connected to node a at a distance of less than 3 hops at time $t_2$; find nodes with a strong direct connection to node a; find nodes directly connected to node a with link type "friend"; find clusters connected to cluster B at time $t_3$.*

ii) where only the connection is specified (i.e. scheme 3: what elements (or subsets) of the set **S** are related in the way **ρ**?)

Find graph objects that are connected in the given way at the given time

$? \, g_1, g_2, t : (g_1, \mathbf{t}) \wedge (g_2, \mathbf{t})$

*e.g. find nodes directly connected with a weight greater than 4 at time 6; find closely connected clusters at time $t_3$.*

(iii) In addition to the tasks constructed according to the pure relational schemes, there is a hybrid variation where we have two graph objects and a given connection relation, and we want to find the *time(s)* at which the objects were connected in the given way:

Find the time points at which two given graph objects were connected in the given way

$? \, t : (\mathbf{g_1}, t) \wedge (\mathbf{g_2}, t)$

*e.g. find the times at which a direct connection exists between graph objects a and b*

**Figure 15 summary of comparison and relation seeking graph structural tasks concerning graph elements, including variations involving specified and unspecified time points**

| | | Graph elements (nodes, graph objects) | | |
|---|---|---|---|---|
| | | **Both constraints** | **One constraint, one target** | **Both are targets** |
| **Time points** | **Both constraints** | **Find connections between elements (comparison)** (How) is graph element $\mathbf{g_1}$ connected to graph element $\mathbf{g_2}$ at the given time, **t**?<br><br>$?\lambda : (\mathbf{g_1}, \mathbf{t}) \, \lambda \, (\mathbf{g_2}, \mathbf{t})$ | **Find elements connected in the given way (relation seeking)** Find the graph element(s) to which graph element $\mathbf{g_1}$ is connected in the given way at time **t**:<br><br>$? \, g_2 : (\mathbf{g_1}, \mathbf{t}) \wedge (g_2, \mathbf{t})$ | **Find elements connected in the given way (relation seeking)** Find graph objects which are connected in the given way at the given time $?g_1, g_2 : (g_1, \mathbf{t}) \wedge (g_2, \mathbf{t})$ |
| | **Both are targets** | **Hybrid** Find the time points at which two given graph objects were connected in the given way<br><br>? $t$ :($\mathbf{g_1}$, $t$)**∧** ($\mathbf{g_2}$, $t$) | **Find elements connected in the given way (relation seeking)** Find the graph element(s) to which graph element $\mathbf{g_1}$ is connected and the time(s) at which the connection(s) occur<br><br>? $g_2$, $t$ : ($\mathbf{g_1}$, $t$) **∧** ($g_2$, $t$) | **Find elements connected in the given way (relation seeking)** Find graph objects (and their associated time points) at any time that are connected in the given way<br><br>? $g_1$, $g_2$, $t$ :($g_1$, $t$)**∧** ($g_2$, $t$) |

#### 4.2.1.3 Compare and find relations between association relations (connections)

The previous tasks deal with finding association relations and finding elements connected in a given way. It is conceivable that we may also wish to compare connections. Comparison of relations does not exist within the original Andrienko framework, and we do not wish to add more task categories than necessary. Therefore, while we use the tasks outlined above to find connection relations, when comparing association relations we treat them as references: either edges in the case of direct connection, or paths (a graph object) in indirect connection. This allows us to use the attribute based tasks of the original Andrienko framework (see section 4.1): direct comparison to compare the attributes associated with the edge or path (e.g. weight, domain attributes, path length), and inverse comparison when we are interested in the (structural) equality of the graph objects themselves. We suggest handling relation seeking tasks involving association relations in a similar manner (e.g. *'find dyads whose edge weight increased between times 3 and 4'* would involve relation seeking where the edge/dyad is the reference and weight is the attribute).

### 4.2.2 Synoptic structural tasks

#### 4.2.2.1 Structural pattern characterisation

Structural pattern characterisation involves describing the pattern associated with one of the four structural behaviours outlined in section 2.2.

#### 4.2.2.2 Structural pattern search

This is the opposite of the above task in that we seek to find the set of graph elements associated with a given pattern or configuration of connections, and/or the time at which they occur.

#### 4.2.2.3 Comparison and relation seeking involving structural patterns

As is the case with the attribute based tasks, we may also wish to compare or find relations between structural patterns, and the graph elements/subsets and time points/intervals associated with these patterns. The many possible permutations of these tasks mirror those of the original attribute based comparison and relation seeking tasks, which are outlined in the task matrices described in section 4.1.

## 4.3 Connection Discovery tasks

We here discuss some examples of the three variations of relational behaviours involved in connection discovery tasks in temporal graphs. We also discuss the possible case of connectional behaviours between graph structural patterns.

### 4.3.1 Heterogeneous behaviours

**(1) Relational behaviour involving two (or more) different attributes of the same reference set**

The formal notation for this behaviour is given in the Andrienko framework as:

(3.43) $\rho(\boldsymbol{f}_1(x), \boldsymbol{f}_2(x) \mid x \in \mathbf{R})$

Where $\boldsymbol{f}_1(x)$ and $\boldsymbol{f}_2(x)$ are two attributes defined on the same reference set **R**.

Applied to temporal graphs, this task considers a relational behaviour between two different attributes and the same graph and temporal components. Note that the Andrienko framework does not explicitly discuss how to handle multiple referrers in relational behaviours. We therefore draw on the partial behaviours to guide us in our discussion.

One example this relational behaviour applied to temporal graphs might be the (partial) relational behaviour between two different attributes of the elements of the graph (nodes or edges) at a single timepoint:

$\rho(\boldsymbol{f}_1(g, t), \boldsymbol{f}_2(g, t) \mid g \in \mathbf{G}, t = \mathbf{t})$

We can imagine representing such a behaviour using a scatterplot, as given in the Andrienko examples. For example, we might consider the relationship between the indegree and out degree of all nodes in the graph at a given time point.

Further, we might consider the relational behaviour between two attribute values for the same graph object over time:

$\rho(\boldsymbol{f}_1(g, t), \boldsymbol{f}_2(g, t) \mid g = \mathbf{g}, t \in \mathbf{T})$

In this case we might look for some correlation or dependency in the two temporal trends, for example, in node indegree and outdegree (e.g. an increase in outdegree followed by an increase in indegree).

Finally, we consider such relations over all graph elements of the time points. The general formula applied to temporal graphs could be written:

$\rho(\boldsymbol{f}_1(g, t), \boldsymbol{f}_2(g, t) \mid g \in \mathbf{G}, t \in \mathbf{T})$

This would consider the relational behaviour between the two attributes for all graph objects at all time points. Perhaps a scatterplot matrix would help us here, or a cumulative view of the temporal trends, depending on the aspect of interest.

**(2) Relational behaviour involving two (or more) different attributes of different reference sets**

The formal notation for relational behaviour involving different attributes of different reference sets is given as

(3.44) $\rho(\boldsymbol{f}_1(x), \boldsymbol{f}_2(z) \mid x \in \mathbf{R}, z \in \mathbf{Z})$

Where $\boldsymbol{f}_1(x)$ is an attribute defined on reference set R, and $\boldsymbol{f}_2$(z) is an attribute defined on a different reference set, **Z**. (However, it is noted that it is highly unlikely that the two reference sets would be completely unrelated.)

We could apply this to temporal graphs in two ways:

*1. Where the reference sets are a graph over time and external events*

In this case, we may wish to investigate the relational behaviour between an attribute of a graph object and external events (in time):

$\rho(\boldsymbol{f}_1(g, t), \boldsymbol{f}_2(z) \mid g \in \mathbf{G}, x_2 \in \mathbf{T}, z \in \mathbf{Z})$

In such a case we may be looking at how the attribute values in the graph are influenced by outside events over time. This may be of particular interest, for example where some form of external intervention in the network is under observation, such as vaccination in a public health network.

*B. Where the reference sets are two different temporal graphs*

Investigate the relations between two (possibly different) attributes of two different graphs over possibly different time periods:

$\rho(\boldsymbol{f}_1(g_1, \mathrm{t}_1), \boldsymbol{f}_2(g_2, t_2) \mid g_1 \in \mathbf{G}_1, t_1 \in \mathbf{T}_1, g_2 \in \mathbf{G}_2, t_2 \in \mathbf{T}_2)$

This behaviour may be of interest where we are investigating two different networks which are related in some way, for example, co-authorship networks from different domains, or the energy grid and a computer network. Moreover, we might not only be interested in attribute based behaviours, but also the *relation between structural patterns*. For example, like Gloor and Zhao [5], we might be interested in the relationship between networks constructed to reflect different communication mediums e.g. face-face, telephone, email. In this case we might also wish to find some correlation between the structural patterns of the network itself, for example, we can imagine that the times at which the email network is densely connected, the face-to-face network may be less so.

### 4.3.2 Homogenous behaviours

**(3) Relational behaviour involving the same attributes of different reference subsets**

The formal notation is given as:

(3.44) $\rho(\boldsymbol{f}(x), \boldsymbol{f}(x') \mid x \in \mathbf{R}_1, x' \in \mathbf{R}_2)$

Applied to temporal graphs, we might consider investigating the relations between different parts of the graph or the graph over different *subsets* of time. We can formulate these to consider whether there is some correlation or influence between attribute values of…

a. different parts of the graph over the same time interval:
$\rho(\boldsymbol{f}(x_1, x_2), \boldsymbol{f}(x'_1, x'_2) \mid x_1 \in \mathbf{O}_1, x_2 \in \mathbf{T}, x'_1 \in \mathbf{O}_2, x'_1 \in \mathbf{T})$

b. the graph or same graph subset during different subsets of time:
$\rho(\boldsymbol{f}(x_1, x_2), \boldsymbol{f}(x'_1, x'_2) \mid x_1 \in \mathbf{O}, x_2 \in \mathbf{T}_1, x'_1 \in \mathbf{O}, x'_1 \in \mathbf{T}_2)$

c. different parts of the graph during different subsets of time:
$\rho(\boldsymbol{f}(x_1, x_2), \boldsymbol{f}(x'_1, x'_2) \mid x_1 \in \mathbf{O}_1, x_2 \in \mathbf{T}_1, x'_1 \in \mathbf{O}_2, x'_1 \in \mathbf{T}_2)$

In the graph case, the role of *graph structure* in relation to attribute values is also very much of interest, for example, we may wish to investigate whether particular structures influence attribute values or vice versa (e.g. Christakis and Fowler's [6] investigation into the influence of network structure on obesity).Moreover we may be interested in relational behaviour between *graph* *structures*. To take a structural example which would reflect formulation (b) above, in social network analysis a number of theories surround tie formation e.g. Yi et al. [7] discuss examples of these including preferential attachment, accumulative advantage (actors with many ties gain more ties), homophily (the theory that those with similar traits connect to one another), follow-the-trend (i.e. the dominant choices of others), and multiconnectivity (a pursuit for diversity and multiplexity). In all cases, we would look for how structural patterns in the graph at one point in time influence the structural patterns at another.

## 4.4 Summary of temporal graph tasks

Our extension to, and instantiation of, the Andrienko task framework has resulted in two main types of temporal graph tasks: structural tasks and attribute-based tasks (note that most attribute-based tasks describe attributes in the context of the graph structure.) We summarise these tasks below:

**Structural tasks include:**

- Find connections between graph elements.
- Find graph elements connected in a given way.
- Comparison and relation seeking between connections.
- Structural pattern characterisation: describe one of four structural patterns associated with given graph elements/subsets and time points/intervals.
- Structural pattern search: find the temporal and graph references associated with one of the above specified patterns.
- Comparison and relation seeking involving structural patterns.
- Connection discovery tasks involving graph structure.

**Attribute based tasks include:**

- Elementary direct lookup: find the value of an attribute of a graph element (node, edge, graph object) at a given time point.
- Elementary inverse lookup: find the graph element(s) and/or time point(s)
- Behaviour characterisation: find the pattern approximating a partial or aspectual behaviour.
- Pattern search: given a pattern describing one of the partial or aspectual behaviours, find the corresponding graph element/subsets and time point/interval over which it occurs.
- Direct comparison: of attribute values associated with a given time point and graph element (elementary) or of partial/aspectual behaviours associated with given time points/intervals and graph elements/subsets (synoptic).
- Inverse comparison: of time points and/or graph elements associated with the occurrence of particular attribute values (elementary), or comparison of time points/intervals and/or graph elements/subsets over which particular patterns of attribute values occur (synoptic).
- Relation seeking: (elementary) find attribute values related in a specified way, possibly with a specified relation on time points and/or graph elements; (synoptic) find patterns related in a specified way, possibly with a specified relation on time intervals and/or graph subsets.
- Connection discovery tasks involving ‘mutual’ behaviours.

Further variations of attribute-based tasks involving the same or different, specified or unspecified temporal and/or graph components are specified in the task matrices.

---

[i] Reduced from: ? $g_2, \lambda$: $\boldsymbol{f}(\mathbf{t}, \mathbf{g}_1) \in \mathbf{C'}$; $\boldsymbol{f}(\mathbf{t}, g_2) \in \mathbf{C''}$; $(\mathbf{t}, \mathbf{g}_1)\ \lambda(\mathbf{t}, g_2)$

[ii] Reduced from: ? $\lambda$: $\boldsymbol{f}(\mathbf{t}_1, \mathbf{g}_1) \in \mathbf{C'}$; $\boldsymbol{f}(\mathbf{t}_2, g_2) \in \mathbf{C''}$; $t_1\ \lambda\ t_2$

[iii] Reduced from ? $t_2, \lambda$: $\boldsymbol{f}(\mathbf{t}_1, \mathbf{g}) \in \mathbf{C'}$; $\boldsymbol{f}(t_2, \mathbf{g}) \in \mathbf{C''}$; $\mathbf{t}_1\ \lambda\ t_2$

[iv] Reduced from ? $t_2, \lambda$: $\boldsymbol{f}(\mathbf{t}_1, \mathbf{g}_1) \in \mathbf{C'}$; $\boldsymbol{f}(t_2, \mathbf{g}_2) \in \mathbf{C''}$; $\mathbf{t}_1\ \lambda\ t_2$

[v] Reduced from ? $t_2, g_2, \lambda$: $\boldsymbol{f}(\boldsymbol{t}_1, \boldsymbol{g}_1) \in \mathbf{C'}$; $\boldsymbol{f}(t_2, g_2) \in \mathbf{C''}$; $(\mathbf{t}_1, \mathbf{g}_1)\ \lambda(t_2, g_2)$

[vi] This is reduced from:? $G', \lambda,$: $\boldsymbol{\beta}(\boldsymbol{f}(x_1, x_2) \mid x_1 \in G',\ x_2 = \mathbf{t}) \approx \mathbf{P}_1$; $\boldsymbol{\beta}(\boldsymbol{f}(x_1, x_2) \mid x_1 \in \mathbf{G''},\ x_2 = \mathbf{t}) \approx \mathbf{P}_2$; $(G', \mathbf{t})\ \lambda\ (\mathbf{G''}, \mathbf{t})$; i.e. all information (the graph subset, timepoint and pattern) is known in the second lookup subtask.